\documentclass[journal]{IEEEtran}
\IEEEoverridecommandlockouts
\usepackage{verbatim}
\usepackage{cite}

\ifCLASSINFOpdf
\usepackage[pdftex]{graphicx}
\usepackage{tikz}
\usepackage{pgfplots}

\pgfplotsset{compat=newest}
\pgfplotsset{plot coordinates/math parser=false}
\usepgfplotslibrary{groupplots}

\newlength\figureheight
\newlength\figureheightd
\newlength\figurewidth
\usepackage[ruled]{algorithm2e} % draw algorithms
\usepackage{amsmath} % improved math formulas
\usepackage{amssymb} % math symbols
\usepackage{amsthm} % theorem environments
\usepackage{thmtools} % extends amsthm with autoref compat. etc.
\usepackage{bm} % bold math
\usepackage{csquotes} % context-sensitive quotation
\usepackage{balance} % balance columns on the final page
\usepackage{hyperref} % hyperlinks via \url

\ifCLASSOPTIONcompsoc
\usepackage[subrefformat=parens,labelformat=parens,ccaption=false,font=normalsize,labelfont=sf,textfont=sf]{subfig}
\else
\usepackage[subrefformat=parens,labelformat=parens,caption=false,font=footnotesize]{subfig}
\fi
\newtheorem{example}{Example}
\newtheorem{theorem}{Theorem}

\newtheorem{definition}{Definition}

\newcommand{\Q}[0]{\mathcal{Q}}
\newcommand{\G}{\mathcal{G}}
\newcommand{\assign}[0]{\bm{P}}
\newcommand{\x}[0]{\bm{x}}
\renewcommand{\c}{\bm{c}}
\newcommand{\y}{\bm{y}}

\newcommand{\Z}{\mathcal{Z}}
\newcommand{\A}[0]{\bm{A}}
\newcommand{\C}[0]{\bm{C}}
\newcommand{\T}[0]{\mathcal{T}}
\newcommand{\V}[0]{\mathcal{V}}
\newcommand{\W}[0]{\mathcal{W}}
\newcommand{\PsiBDC}{\bm{\Psi}_\mathsf{BDC}}
\newcommand{\PsiBDCLT}{\bm{\Psi}_\mathsf{BDC-LT}}
\newcommand{\PsiLT}{\bm{\Psi}_\mathsf{LT}}

\newcommand{\me}{\mathrm{e}}
\newcommand{\var}{\mathop{\rm{Var}}}
\newcommand{\E}{\mathop{\mathbb{E}}}
\newcommand*\dif{\mathop{}\!\mathrm{d}}

\newcommand{\fexp}{L_{\text{BDC}}}
\newcommand{\enc}{\bm{\Psi}}

\newcommand{\cenc}{\sigma_{\mathsf{encode}}}
\newcommand{\cencbdc}{\sigma_{\mathsf{encode, BDC}}}
\newcommand{\cenclt}{\sigma_{\mathsf{encode, LT}}}
\newcommand{\cmap}{\sigma_{\mathsf{map}}}
\newcommand{\cred}{\sigma_{\mathsf{reduce}}}
\newcommand{\credbdc}{\sigma_{\mathsf{reduce, BDC}}}

\newcommand{\ca}{\sigma_{\mathsf{A}}}
\newcommand{\cm}{\sigma_{\mathsf{M}}}

\newcommand{\tenc}{D_{\mathsf{encode}}}
\newcommand{\tmap}{D_{\mathsf{map}}}

\newcommand{\tred}{D_{\mathsf{reduce}}}

\newcommand{\Pf}{P_\mathsf{f}}
\newcommand{\Pft}{P_\mathsf{f, target}}

\begin{document}

\title{Block-Diagonal and LT Codes for Distributed Computing With Straggling
  Servers}

\author{
  \IEEEauthorblockN{
    Albin Severinson, \emph{Student Member, IEEE},
    Alexandre Graell i Amat, \emph{Senior Member, IEEE}, \\
    and Eirik Rosnes, \emph{Senior Member, IEEE}
  }

  \thanks{This work was presented in part at the IEEE Information
    Theory Workshop (ITW), Kaohsiung, Taiwan, November 2017, and it
    was published in part in ``Coding for Distributed Computing:
    Investigating and improving upon coding theoretical frameworks for
    distributed computing,'' MSc. thesis, Chalmers University of
    Technology, June 2017. This work was funded by the Swedish
    Research Council under grant 2016-04253 and the Research Council
    of Norway under grant 240985/F20.

    Albin Severinson was with the Department of Electrical Engineering, Chalmers
    University of Technology, SE-41296 Gothenburg, Sweden. He is now with
    Simula UiB and the Department of Informatics at the University of Bergen,
    N-5020 Bergen, Norway (email: albin@severinson.org).

    Alexandre Graell i Amat is with the Department of Electrical Engineering,
    Chalmers University of Technology, SE-41296 Gothenburg, Sweden (email:
    alexandre.graell@chalmers.se).

    Eirik Rosnes is with Simula UiB, N-5020 Bergen, Norway (email:
    eirikrosnes@simula.no).

  } }

\maketitle

\begin{abstract}
  We propose two coded schemes for the distributed computing problem
  of multiplying a matrix by a set of vectors. The first scheme is
  based on partitioning the matrix into submatrices and applying
  maximum distance separable (MDS) codes to each submatrix. For this
  scheme, we prove that up to a given number of partitions the
  communication load and the computational delay (not including the
  encoding and decoding delay) are identical to those of the scheme
  recently proposed by Li \emph{et al.}, based on a single, long MDS
  code. However, due to the use of shorter MDS codes, our scheme
  yields a significantly lower overall computational delay when the
  delay incurred by encoding and decoding is also considered. We
  further propose a second coded scheme based on Luby Transform (LT)
  codes under inactivation decoding.  Interestingly, LT codes may
  reduce the delay over the partitioned scheme at the expense of an
  increased communication load. We also consider distributed computing
  under a deadline and show numerically that the proposed schemes
  outperform other schemes in the literature, with the LT code-based
  scheme yielding the best performance for the scenarios considered.
\end{abstract}
\IEEEpeerreviewmaketitle


\section{Introduction}

Distributed computing systems have emerged as one of the most
effective ways of solving increasingly complex computational problems,
such as those in large-scale machine learning and data analytics
\cite{Barroso2009,Chen2014,Barroso2011}. These systems, referred to as
``warehouse-scale computers'' (WSCs) \cite{Barroso2009}, may be
composed of thousands of relatively homogeneous hardware and software
components. Achieving high availability and efficiency for
applications running on WSCs is a major challenge. One of the main
reasons is the large number of components that may experience
transient or permanent failures \cite{Barroso2011}. As a result,
several distributed computing frameworks have been proposed
\cite{Dean2004,Zaharia2016,Ranjan2014}. In particular, MapReduce
\cite{Dean2004} has gained significant attention as a means of
effectively utilizing large computing clusters. For example, Google
routinely performs computations over several thousands of servers
using MapReduce \cite{Dean2004}. Among the challenges brought on by
distributed computing systems, the problems of straggling servers and
bandwidth scarcity have recently received significant attention. The
straggler problem is a synchronization problem characterized by the
fact that a distributed computing task must wait for the slowest
server to complete its computation, which may cause large delays
\cite{Dean2004}. On the other hand, distributed computing tasks
typically require that data is moved between servers during the
computation, the so-called \textit{data shuffling}, which is a
challenge in bandwidth-constrained networks.

Coding for distributed computing to reduce the computational delay and
the communication load between servers has recently been considered in
\cite{Li2015,Lee2017}. In \cite{Li2015}, a structure of repeated
computation tasks across servers was proposed, enabling coded
multicast opportunities that significantly reduce the required
bandwidth to shuffle the results. In \cite{Lee2017}, the authors
showed that maximum distance separable (MDS) codes can be applied to a
linear computation task (e.g., multiplying a vector with a matrix) to
alleviate the effects of straggling servers and reduce the
computational delay. In \cite{Li2016}, a unified coding framework was
presented and a fundamental tradeoff between computational delay and
communication load was identified. The ideas of \cite{Li2015,Lee2017}
can be seen as particular instances of the framework in \cite{Li2016},
corresponding to the minimization of the communication load and the
computational delay, respectively. The code proposed in \cite{Li2016} is an MDS code
of code length proportional to the number of rows of the matrix to be
multiplied, which may be very large in practice. For example, Google
performs matrix-vector multiplications with matrices of dimension of
the order of $10^{10}\times 10^{10}$ when ranking the importance of
websites \cite{Ishii2014}. In \cite{Li2015,Lee2017,Li2016}, the
computational delay incurred by the encoding and decoding is not
considered. However, the encoding and decoding may incur a high
computational delay for large matrices.

Coding has previously been applied to several related problems in
distributed computing. For example, the scheme in \cite{Lee2017} has
been extended to distributed matrix-matrix multiplication where both
matrices are too large to be stored at one server
\cite{Lee2017a,Qian2017}. Whereas the schemes in
\cite{Lee2017,Lee2017a} are based on MDS codes, the scheme in
\cite{Qian2017} is based on a novel coding scheme that exploits the
algebraic properties of matrix-matrix multiplication over a finite
field to reduce the computational delay. In \cite{Dutta2016}, it was
shown that introducing sparsity in a structured manner during encoding
can speed up computing dot products between long vectors. Distributed
computing over heterogeneous clusters has been considered in
\cite{Reisizadeh2017}.

In this paper, we propose two coding schemes for the problem of
multiplying a matrix by a set of vectors. The first is a
block-diagonal coding (BDC) scheme equivalent to partitioning the
matrix and applying smaller MDS codes to each submatrix separately (we
originally introduced the BDC scheme in \cite{Severinson2017}). The
storage design for the BDC scheme can be cast as an integer
optimization problem, whose computation scales exponentially with the
problem size. We propose a heuristic solver for efficiently solving
the optimization problem, and a branch-and-bound approach for
improving on the resulting solution iteratively. Furthermore, we prove
that up to a certain level of partitioning the BDC scheme has
identical computational delay (as defined in \cite{Li2016}) and
communication load to those of the scheme in
\cite{Li2016}. Interestingly, when the delay incurred by encoding and
decoding is taken into account, the proposed scheme achieves an
overall computational delay significantly lower than that of the
scheme in \cite{Li2016}. We further propose a second coding scheme
based on Luby Transform (LT) codes \cite{Luby2002} under inactivation
decoding \cite{rfc6330}, which in some scenarios achieves a lower
computational delay than that of the BDC scheme at the expense of a
higher communication load. We show that for the LT code-based scheme
it is possible to trade an increase in communication load for a lower
computational delay. We finally consider distributed computing under a
deadline, where we are interested in completing a computation within
some computational delay, and show numerically that both the BDC and
the LT code-based schemes significantly increase the probability of
meeting a deadline over the scheme in \cite{Li2016}. In particular,
the LT code-based scheme achieves the highest probability of meeting a
deadline for the scenarios considered.

\section{System Model and Preliminaries}
\label{sec:SystemModel}

We consider the distributed matrix multiplication problem, i.e., the
problem of multiplying a set of vectors with a matrix. In particular,
given an $m \times n$ matrix $\A \in \mathbb{F}^{m \times n}_{2^l}$
and $N$ vectors $\x_1, \ldots, \x_N \in \mathbb{F}^{n}_{2^l}$, where
$\mathbb{F}_{2^l}$ is an extension field of characteristic $2$, we
want to compute the $N$ vectors
$\bm{y}_1 = \bm{Ax}_1, \ldots, \bm{y}_N = \bm{Ax}_N$. The computation
is performed in a distributed fashion using $K$ servers,
$S_1, \ldots, S_K$. Each server is responsible for multiplying $\eta m$
matrix rows by the vectors $\x_1, \ldots, \x_N$, for some
$\frac{1}{K} \leq \eta \leq 1$. We refer to $\eta$ as the fraction of
rows stored at each server and we assume that $\eta$ is selected such
that $\eta m$ is an integer. Prior to computing
$\bm{y}_1, \ldots, \bm{y}_N$, $\A$ is encoded by an $r\times m$
encoding matrix $\bm{\Psi}=[\Psi_{i,j}]$, resulting in the coded
matrix $\bm{C} = \bm{\Psi \A}$, of size $r\times n$, i.e., the rows of
$\A$ are encoded using an $(r,m)$ linear code with $r \geq m$.  This
encoding is carried out in a distributed manner over the $K$ servers
and is used to alleviate the straggler problem. We allow assigning
each row of the coded matrix $\bm{C}$ to several servers to enable
coded multicasting, a strategy used to address the bandwidth scarcity
problem. Let
\begin{equation}
  \notag
  q=K\frac{m}{r},
\end{equation}
where we assume that $r$ divides $Km$ and hence $q$ is an
integer. The $r$ coded rows of $\C$, $\c_1,\ldots, \c_r$, are divided into
$\binom{K}{\eta q}$ disjoint batches, each containing $r/ \binom{K}{\eta q}$ coded
rows. Each batch is assigned to $\eta q$ servers. Correspondingly, a batch $B$ is
labeled by a unique set $\T \subset \{S_1, \ldots, S_K\}$, of size $|\T|=\eta q$,
denoting the subset of servers that store that batch. We write $B_\T$ to denote
the batch stored at the unique set of servers $\T$. Server $S_k$, $k=1, \ldots,
K$, stores the coded rows of $B_\T$ if and only if $S_k \in \T$.

\subsection{Probabilistic Runtime Model}
\label{sec:ProbRuntime}

We assume that running a computation on a single server takes a random
amount of time, which is denoted by the random variable $H$, according
to the shifted-exponential cumulative probability distribution
function (CDF)
\begin{equation} \notag
  F_H(h; \sigma) = \begin{cases}
    1 - \me^{-\left(\frac{h}{\sigma} - 1\right)}, & \text{for $h \geq \sigma$} \\
    0, & \text{otherwise}
  \end{cases},
\end{equation}
where $\sigma$ is a parameter used to scale the distribution. Denote
by $\ca$ and $\cm$ the number of time units required to complete one
addition and one multiplication (over $\mathbb{F}_{2^l}$),
respectively, over a single server. Let $\sigma$ be the weighted sum
of the number of additions and multiplications required to complete
the computation, where the weighting coefficients are $\ca$ and $\cm$,
respectively. As in \cite{Edmonds2017}, we assume that $\ca$ is in
$\mathcal{O}(\frac{l}{64})$ and $\cm$ in $\mathcal{O}(l \log_2
l)$. Furthermore, we assume that the hidden coefficients are
comparable and will thus not consider them. With some abuse of
language, we refer to the parameter $\sigma$ associated with some
computation as its computational complexity. For example, the
complexity (number of time units) of computing the inner product of
two length-$n$ vectors is $\sigma = (n-1)\ca + n \cm$ as it requires
performing $n-1$ additions and $n$ multiplications. The shift of the
shifted-exponential distribution should be interpreted as the minimum
amount of time the computation can be completed in. The tail of the
distribution accounts for transient disturbances that are at the root
of the straggler problem. These include transmission and queuing
delays during initialization as well as contention for the local disk
and slow-downs due to higher priority tasks being assigned to the same
server \cite{Verma2015}. The complexity of a computation $\sigma$
affects both the shift and the tail of the distribution since the
probability of transient behavior occurring increases with the amount
of time the computation is running.  In the results section we also
consider a model where $\sigma$ only affects the shift. The
shifted-exponential distribution was proposed as a model for the
latency of file queries from cloud storage systems in \cite{Liang2014}
and was subsequently used to model computational delay in
\cite{Lee2017,Li2016}.

When an algorithm is split into $K$ parallel subtasks that are run
across $K$ servers, we denote the runtime of the subtask running on
server $S_k$ by $H_k$.  As in \cite{Lee2017}, we assume that
$H_1, \ldots, H_K$ are independent and identically distributed random
variables with CDF $F_{H}(K h;\sigma)$. For $i=1, \dots, K$, we
denote the $i$-th order statistic by $H_{(i)}$, i.e., the $i$-th
smallest random variable of $H_{1}, \ldots, H_{K}$. The runtime of the
$i$-th fastest server to complete its subtask is thus given by
$H_{(i)}$, which is a Gamma distributed random variable with
expectation and variance given by \cite{Arnold2008}

\makeatletter
\if@twocolumn
\begin{equation} \notag
  \mu(\sigma, K, i) \triangleq \E \left[ H_{(i)} \right]
  = \sigma \left(1 + \sum_{j=K-i+1}^K \frac{1}{j} \right),
\end{equation}

\begin{equation}
  \notag
  \var \left[H_{(i)} \right] = \sigma^2 \sum_{j=K-i+1}^K \frac{1}{j^2}.
\end{equation}
\else
\begin{equation} \notag
  \mu(\sigma, K, g) \triangleq \mathop{\mathbb{E}} \left(H_{(i)} \right) =
  \sigma \left(1 + \sum_{j=K-i+1}^K \frac{1}{j} \right),\;
  \mathop{\rm{Var}} \left(H_{(i)} \right) = \sigma^2 \sum_{j=K-i+1}^K \frac{1}{j^2}.
\end{equation}
\fi
\makeatother

%For a system with $K$ servers performing a computation of complexity
%$\sigma$ we denote by $\mu(\sigma, K, i)$ the expected runtime of the
%$i$-th fastest server $H_{(i)}$.
We parameterize the Gamma
distribution by its inverse scale factor $a$ and its shape parameter
$b$. We give these in terms of the distribution mean and variance as
\cite{Walck2007}

\begin{equation}
  \notag
  a = \frac{\E [H_{(i)} ] - \sigma}{\var [H_{(i)}]}
  \;\text{ and }\;
  b = \frac{\left(\E [H_{(i)} ] - \sigma\right)^2}{\var [H_{(i)}]}.
\end{equation}
Denote by $F_{H_{(i)}}(h_{(i)}; \sigma, K)$ the CDF of $H_{(i)}$. It is
given by \cite{Walck2007}
\begin{equation} \notag
  F_{H_{(i)}}(h_{(i)}; \sigma, K) = \begin{cases}
    \frac{\gamma(b, a(h_{(i)}-\sigma))}{\Gamma(b)}, & \text{for $h_{(i)} \geq \sigma$} \\
    0, & \text{otherwise}\\
  \end{cases},
\end{equation}
where $\Gamma$ denotes the Gamma function and $\gamma$ the lower
incomplete Gamma function,
\begin{align*}
  \notag
  \Gamma(b) = \int_0^{\infty} x^{b-1} \me^{-x} \dif x
  \;\text{ and }\;
  \gamma(b, ah) = \int_0^{ah} x^{b-1} \me^{-x} \dif x.
\end{align*}
We remark that $F_{H_{(i)}}(h_{(i)}; \sigma, K)$ is the probability of
a computation finishing prior to some deadline $t=h_{(i)}$.

\subsection{Distributed Computing Model}
\label{sec:computing_model}

We consider the coded computing framework introduced in \cite{Li2016},
which extends the MapReduce framework \cite{Dean2004}. The overall
computation proceeds in three phases, the \textit{map},
\textit{shuffle}, and \textit{reduce} phases, which are augmented to
make use of the coded multicasting strategy proposed in \cite{Li2015}
to address the bandwidth scarcity problem and the coded scheme
proposed in \cite{Lee2017} to alleviate the straggler
problem. Furthermore, we consider the delay incurred by the encoding
of $\A$ that takes place before the start of the map phase. We refer
to this as the \textit{encoding} phase. Also, we assume that the
matrices $\A$ and $\enc$ as well as the input vectors
$\x_1, \ldots, \x_N$ are known to all servers at the start of the
computation. The overall computation proceeds in the following manner.

\subsubsection{Encoding Phase}
In the encoding phase, the coded matrix $\C$ is computed from $\A$ and
$\enc$ in a distributed fashion. Specifically, denote by
$\mathcal{R}^{(S)}$ the set of indices of rows of $\C$ that are
assigned to server $S$ and denote by $\enc^{(S)}$ the matrix
consisting of the rows of $\enc$ with indices from
$\mathcal{R}^{(S)}$. Then, server $S$ computes the coded rows it needs
by multiplying $\enc^{(S)}$ by $\A$.  Note that since we assign each
coded row to $\eta q$ servers, each row of $\C$ is computed separately
by $\eta q$ servers. We define the computational delay of the encoding
phase as its average runtime per source row and vector $\y$, i.e.,
\begin{equation}
  \notag
  \tenc = \frac{\eta q}{mN} \mu\left(\frac{\cenc}{K}, K, K\right),
\end{equation}
where $\cenc$ is the complexity of the encoding. During the encoding
process, the rows of $\enc$ are multiplied by the columns of $\A$.
Therefore, the complexity scales with the product of the number of
nonzero elements of $\enc$ and the number of columns of $\A$.
Specifically,
  \begin{equation}
    \notag
  \cenc = \left\vert \{(i, j) : \Psi_{i, j} \neq 0 \} \right\vert n
  \left(\ca + \cm\right) - n\ca.
\end{equation}

Alternatively, we compute $\C$ by performing a decoding operation on
$\A$. In this case $\cenc$ is the decoding complexity (see
Section~\ref{computational_delay}). Furthermore, since the decoding
algorithms are designed to decode the entire codeword, each server has
to compute all rows of $\C$. Using this strategy the encoding delay is
\begin{equation} \notag
  \tenc = \frac{K}{mN} \mu\left(\frac{\cenc}{K}, K, K\right).
\end{equation}
For each case we choose the strategy that minimizes the delay.

\subsubsection{Map Phase}

\begin{figure}[t]
  \centering
  \resizebox{1\columnwidth}{!}{
    \includegraphics[width=1\linewidth]{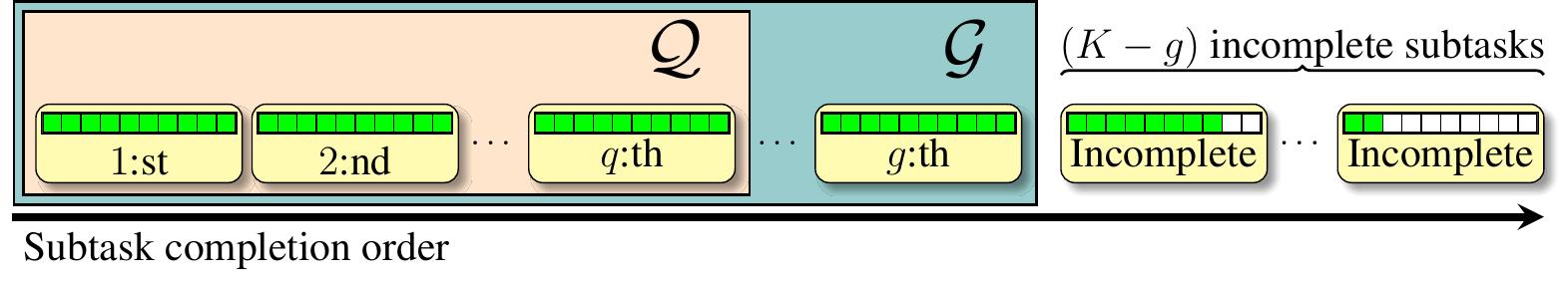}
  }
  \vspace{-3ex}
  \caption{Servers (yellow boxes) finish their respective subtasks in random order.}
  \label{fig:subtask_order}
  \vspace{-2ex}
\end{figure}

In the map phase, we compute in a distributed fashion coded intermediate values,
which will be later used to obtain vectors $\y_1,\ldots,\y_N$. Server $S$
multiplies the input vectors $\x_j$, $j=1,\ldots,N$, by all the coded rows of
matrix $\C$ it stores, i.e., it computes
\begin{equation} \notag
  \Z_{j}^{(S)} = \{\bm{c} \x_j :\; \c\in \{B_\T :\; S \in \T\}\},\; j=1, \ldots, N.
\end{equation}

The map phase terminates when a set of servers
$\G \subseteq \{S_1, \ldots, S_K\}$ that collectively store enough
values to decode the output vectors have finished their map
computations. We denote the cardinality of $\G$ by $g$. The $(r,m)$
linear code proposed in \cite{Li2016} is an MDS code for which
$\y_1,\ldots,\y_N$ can be obtained from any subset of $q$ servers,
i.e., $g = q$. We illustrate the completion of subtasks in
Fig.~\ref{fig:subtask_order}.

We define the computational delay of the map phase as its average
runtime per source row and vector $\y$, i.e.,
\begin{equation}
  \notag
  \tmap = \frac{1}{mN} \mu\left(\frac{\cmap}{K}, K, g\right),
\end{equation}
where $\cmap = K \eta m N \left( (n-1)\ca + n \cm \right)$, as all $K$
servers compute $\eta m$ inner products, each requiring $n-1$ additions
and $n$ multiplications, for each of the $N$ input vectors. In
\cite{Li2016}, $\tmap$ is referred to simply as the computational
delay.

After the map phase, the computation of $\y_1,\ldots,\y_N$ proceeds
using only the servers in $\G$. We denote by $\Q \subseteq \G$ the set
of the first $q$ servers to complete the map phase. Each of the $q$
servers in $\Q$ is responsible to compute $N/q$ of the vectors
$\y_1,\ldots,\y_N$. Let $\W_S$ be the set containing the indices of
the vectors $\y_1,\ldots,\y_N$ that server $S\in\Q$ is responsible
for. The remaining servers in $\G$ assist the servers in $\Q$ in the
shuffle phase.

\subsubsection{Shuffle Phase}
\label{sec:shuffle_phase}

In the shuffle phase, intermediate values calculated in the map phase
are exchanged between servers in $\G$ until all servers in $\Q$ hold
enough values to compute the vectors they are responsible for. As in
\cite{Li2016}, we allow creating and multicasting coded messages that
are simultaneously useful for multiple servers. Furthermore, as in
\cite{Lee2017}, we denote by $\phi(j)$ the ratio between the
communication load of unicasting the same message to each of $j$
recipients and multicasting that message to $j$ recipients.  For
example, if the communication load of multicasting a message to $j$
recipients and unicasting a message to a single recipient is the same,
we have $\phi(j) = j$. On the other hand, if the communication load of
multicasting a message to $j$ recipients is equal to that of
unicasting that same message to each recipient, $\phi(j) = 1$.  The
total communication load of a multicast message is then given by
$\frac{j}{\phi(j)}$. The shuffle phase proceeds in three steps as
follows.
\begin{enumerate}
\item Coded messages composed of several intermediate values are
  multicasted among the servers in $\Q$.
\item Intermediate values are unicasted among the servers in $\Q$.
\item Any intermediate values still missing from servers in $\Q$ are
  unicasted from the remaining servers in $\G$, i.e., from the servers
  in $\G \setminus \Q$.
\end{enumerate}

For a subset of servers $\mathcal{S} \subset \Q$ and
$S \in \Q \setminus \mathcal{S}$, we denote the set of intermediate
values needed by server $S$ and known \textit{exclusively} by the
servers in $\mathcal{S}$ by $\V_{\mathcal{S}}^{(S)}$. More formally,
\begin{equation}
  \notag
  \V_{\mathcal{S}}^{(S)} \triangleq
  \{ \bm{c} \x_j :\; j\in \W_S
  \text{ and }
  \c \in \{B_\T :\; \T \cap \Q =\mathcal{S} \} \}.
\end{equation}

We transmit coded multicasts only between the servers in $\Q$, and
each coded message is simultaneously sent to multiple servers. We
denote by
\begin{equation}
  \label{eq:sq}
  s_q \triangleq \text{inf}\left(s : \sum_{j=s}^{\eta q} \alpha_j \leq
    1 - \eta \right),\;
  \alpha_j \triangleq \frac{\binom{q-1}{j} \binom{K-q}{\eta q -
    j}}{\frac{q}{K} \binom{K}{\eta q}},
\end{equation}
the smallest number of recipients of a coded message \cite{Li2016}. We
remark that $m \alpha_j$ is the total number of coded values delivered
to each server via the coded multicast messages with exactly $j$
recipients. More specifically, for each
$j \in \{ \eta q, \eta q - 1, \ldots, s_q\}$, and every subset
$\mathcal{S} \subseteq \Q$ of size $j + 1$, the shuffle phase proceeds
as follows.

\begin{enumerate}
\item For each $S \in \mathcal{S}$, we evenly and arbitrarily split
  $\V_{\mathcal{S} \setminus S}^{(S)}$ into $j$ disjoint segments,
  $\V_{\mathcal{S} \setminus S}^{(S)} = \{ \V_{\mathcal{S} \setminus
    S, \tilde{S}}^{(S)} : \tilde{S} \in \mathcal{S} \setminus S \}$,
  and associate the segment
  $\V_{\mathcal{S} \setminus S, \tilde{S}}^{(S)}$ to server
  $\tilde{S}$.

\item Server $\tilde{S} \in \mathcal{S}$ multicasts the bit-wise
  modulo-$2$ sum of all the segments associated to it in
  $\mathcal{S}$. More precisely, it multicasts
  $\mathop{\oplus}_{S \in \mathcal{S} \setminus \tilde{S}}
  \V_{\mathcal{S} \setminus S,\tilde{S}}^{(S)}$ to the other servers
  in $\mathcal{S} \setminus \tilde{S}$, where $\mathop{\oplus}$
  denotes the modulo-$2$ sum operator.
\end{enumerate}

By construction, exactly one value that each coded message is composed
of is unknown to each recipient. The other values have been computed
locally by the recipient. More precisely, for every pair of servers
$S,\tilde{S}\in\mathcal{S}$, since server $S$ has computed locally the
segments $\V_{\mathcal{S} \setminus S', \tilde{S}}^{(S')}$ for all
$S' \in \mathcal{S} \setminus \{\tilde{S}, S\}$, it can cancel them
from the message sent by server $\tilde{S}$, and recover the intended
segment. We finish the shuffle phase by either unicasting any
remaining needed values until all servers in $\Q$ hold enough
intermediate values to decode successfully, or by repeating the above
two steps for $j=s_q - 1$. We refer to these alternatives as shuffling
strategy $1$ and $2$, respectively. We always select the strategy
achieving the lowest communication load. If any server in $\Q$ still
needs more intermediate values at this point, they are unicasted from
other servers in $\G$. This may happen only if a non-MDS code is
used. We remark that it may be possible to opportunistically create
additional coded multicasting opportunities by exploiting the
remaining $g-q$ servers in $\G$.

\begin{definition}
  \label{def:L}
  The \textit{communication load}, denoted by $L$, is the number of
  unicasts and multicasts (weighted by their cost relative to a
  unicast) per source row and vector $\y$ exchanged during the shuffle
  phase. Specifically, each unicasted message increases $L$ by
  $\frac{1}{mN}$, and each message multicasted to $j$ recipients increases $L$ by
  $\frac{j}{mN \phi(j)}$.
\end{definition}

The communication load after completing the shuffle phase is given in
\cite{Li2016}. If the shuffle phase finishes by unicasting the
remaining needed values (strategy $1$), the communication load after
completing the multicast phase is
\begin{equation}
  \notag
  \sum_{j=s_q}^{\eta q} \frac{\alpha_j}{\phi(j)}.
\end{equation}
If instead steps $1)$ and $2)$ are repeated for $j=s_q - 1$ (strategy
$2$), the communication load is
\begin{equation}
  \notag
  \sum_{j=s_q-1}^{\eta q} \frac{\alpha_j}{\phi(j)}.
\end{equation}
For the scheme in \cite{Li2016}, the total communication load is
\begin{equation}
  \label{eq:unpartitioned_load}
  L_{\mathsf{MDS}} = \min\left( \sum_{j=s_q}^{\eta q}
    \frac{\alpha_j}{\phi(j)} + 1 - \eta - \sum_{j=s_q}^{\eta q} \alpha_j, \sum_{j=s_q -1 }^{\eta q}
    \frac{\alpha_j}{\phi(j)} \right),
\end{equation}
where $1 - \eta - \sum_{j=s_q}^{\eta q} \alpha_j$ is the communication
load due to unicasting the remaining needed values.

\subsubsection{Reduce Phase}
Finally, in the reduce phase, the vectors $\y_1,\ldots,\y_N$ are
computed. More specifically, server $S\in \Q$ uses the locally
computed sets $\Z_{1}^{(S)}, \ldots, \Z_{N}^{(S)}$ and the received
messages to compute the vectors $\y_j$, $\forall j \in \W_{S}$. The
computational delay of the reduce phase is its average runtime per
source row and output vector $\y$, i.e.,
\begin{equation}
  \notag
  \tred = \frac{1}{mN} \mu\left(\frac{\cred}{q}, q, q\right),
\end{equation}
where $\cred$ is the computational complexity (see
Section~\ref{sec:ProbRuntime}) of the reduce phase.

\begin{definition}
\label{def:D}
The \textit{overall computational delay}, $D$, is the sum of the
encoding, map, and reduce phase delays, i.e.,
$D = \tenc + \tmap + \tred$.
\end{definition}

\subsection{Previously Proposed Coded Computing Schemes}
\label{sec:schemes}

Here we formally define the uncoded scheme (UC) and the coded
computing schemes of \cite{Li2015,Lee2017,Li2016} (which we refer to
as the straggler coding (SC), coded MapReduce (CMR), and unified
scheme, respectively) in terms of the model above. Specifically, to
make a fair comparison with our coded computing scheme with parameters
$K$, $q$, $m$, and $\eta$, we define the corresponding uncoded, CMR,
SC, and unified schemes. When referring to the system parameters of a
given scheme, we will write the scheme acronym in the subscript. We
only explicitly mention the parameters that differ. The number of
servers $K$ is unchanged for all schemes considered.

The uncoded scheme uses no erasure coding and no coded multicasting
and has parameters $\eta_{\mathsf{UC}} = \frac{1}{K}$ and
$q_{\mathsf{UC}} = K$, implying
$\eta_{\mathsf{UC}} q_{\mathsf{UC}} = 1$.  Furthermore, the encoding
matrix $\enc_{\mathsf{UC}}$ is the $m \times m$ identity matrix and
the coded matrix is $\C_{\mathsf{UC}} = \A$.

The CMR scheme \cite{Li2015} uses only coded multicasting, i.e., $\C_{\mathsf{CMR}}
= \A$ and $q_{\mathsf{CMR}} = K$. Furthermore, the fraction of rows stored at each
server is $\eta_{\mathsf{CMR}} = \frac{\eta q}{K}$. We remark that there is no reduce
delay for this scheme, i.e., $\tred = 0$.

The SC scheme \cite{Lee2017} uses an erasure code but no coded
multicasting. For the corresponding SC scheme, the code rate is
unchanged, i.e., $q_{\mathsf{SC}} = q$, and the fraction of rows
stored at each server is
$\eta_{\mathsf{SC}} = \frac{1}{q_{\mathsf{SC}}}$. The encoding matrix
$\enc_{\mathsf{SC}}$ of the SC scheme is obtained by splitting the
rows of $\A$ into $q_{\rm{SC}}$ equally tall submatrices
$\A_1, \ldots, \A_{q_{\mathsf{SC}}}$ and applying a
$(K, q_{\mathsf{SC}})$ MDS code to the elements of each submatrix,
thereby creating $K$ coded submatrices $\C_1, \ldots, \C_K$. The coded
matrix $\C_{\mathsf{SC}}$ is the concatenation of
$\C_1, \ldots, \C_K$, i.e.,
\begin{equation}
  \notag
  \C_{\mathsf{SC}} =
  \left(\begin{matrix}
    \C_1 \\
    \vdots \\
    \C_K
  \end{matrix}\right).
\end{equation}

The unified scheme \cite{Li2016} uses both an erasure code and coded
multicasting and has parameters $\eta_{\mathsf{unified}}=\eta$ and
$q_{\mathsf{unified}}=q$.  Furthermore, the encoding matrix of the
unified scheme, $\enc_{\mathsf{unified}}$, is an $(r, m)$ MDS code
encoding matrix.

\section{Block-Diagonal Coding}
\label{sec:bdc}

\begin{figure}[t]
  \centering
  \resizebox{1\columnwidth}{!}{
    \includegraphics{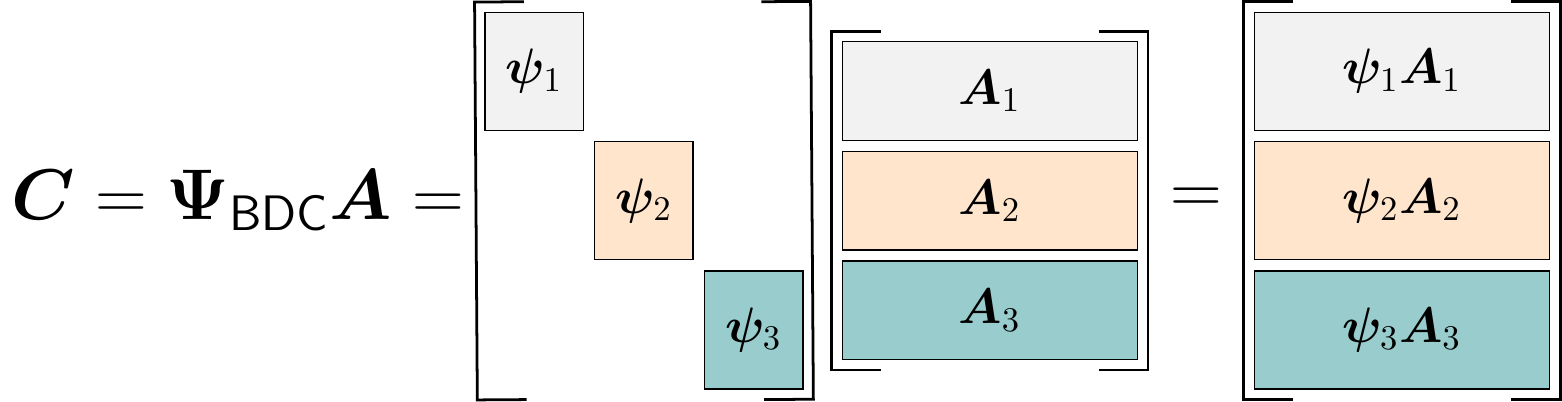}
  }
  \vspace{-2ex}
  \caption{BDC scheme with $T=3$ partitions.}
  \label{fig:bdc_example}
  \vspace{-2ex}
\end{figure}

In this section, we introduce a BDC scheme for the problem of
multiplying a matrix by a set of vectors. For large matrices, the
encoding and decoding complexity of the proposed scheme is
significantly lower than that of the scheme in \cite{Li2016}, leading
to a lower overall computational delay, as will be shown in
Section~\ref{sec:results}.  Specifically, the scheme is based on a block-diagonal encoding
matrix of the form
\begin{equation} \notag
  \PsiBDC =
  \begin{bmatrix}
    \bm{\psi}_1 & &  \\
    & \ddots & \\
    & & \bm{\psi}_T
  \end{bmatrix},
\end{equation}
where $\bm{\psi}_1, \ldots, \bm{\psi}_T$ are
$\frac{r}{T}\times \frac{m}{T}$ encoding matrices of an
$(\frac{r}{T},\frac{m}{T})$ MDS code, for some integer $T$ that
divides $m$ and $r$. Note that the encoding given by $\PsiBDC$ amounts
to partitioning the rows of $\A$ into $T$ disjoint submatrices
$\A_1, \ldots, \A_T$ and encoding each submatrix separately. We refer
to an encoding $\PsiBDC$ with $T$ disjoint submatrices as a
$T$-partitioned scheme, and to the submatrix of $\C=\PsiBDC\A$
corresponding to $\bm{\psi}_i$ as the $i$-th partition. We remark that
all submatrices can be encoded using the same encoding matrix, i.e.,
$\bm{\psi}_i=\bm{\psi}$, $i=1,\ldots,T$, reducing the storage
requirements, and encoding/decoding can be performed in parallel if
many servers are available. Notably, by keeping the ratio
$\frac{m}{T}$ constant, the decoding complexity scales linearly with
$m$.  We further remark that the case $\PsiBDC=\bm{\psi}$ (i.e., the
number of partitions is $T=1$) corresponds to the scheme in
\cite{Li2016}, which we will sometimes refer to as the
\emph{unpartitioned} scheme. We illustrate the BDC scheme with $T=3$
partitions in Fig.~\ref{fig:bdc_example}.

\subsection{Assignment of Coded Rows to Batches}
\label{sec:assignment}

For a block-diagonal encoding matrix $\PsiBDC$, we denote by
$\bm{c}_i^{(t)}$, $t=1,\ldots,T$ and $i=1,\ldots,r/T$, the $i$-th
coded row of $\C$ within partition $t$. For example, $\bm{c}_1^{(2)}$
denotes the first coded row of the second partition. As described in
Section~\ref{sec:SystemModel}, the coded rows are divided into
$\binom{K}{\eta q}$ disjoint batches. To formally describe the
assignment of coded rows to batches we use a
$\binom{K}{\eta q} \times T$ integer matrix $\assign=[p_{i,j}]$, where
$p_{i,j}$ is the number of rows from partition $j$ that are stored in
batch $i$. In the sequel, $\assign$ will be referred to as the
assignment matrix.  Note that, due to the MDS property, any set of
$m/T$ rows of a partition is sufficient to decode the partition. Thus,
without loss of generality, we consider a \emph{sequential} assignment
of rows of a partition into batches. More precisely, when first
assigning a row of partition $t$ to a batch, we pick
$\bm{c}_1^{(t)}$. Next time a row of partition $t$ is assigned to a
batch we pick $\bm{c}_2^{(t)}$, and so on. In this manner, each coded
row is assigned to a unique batch exactly once. The rows of $\assign$
are labeled by the subset of servers the corresponding batch is stored
at, and the columns are labeled by their partition indices. For
convenience, we refer to the pair $(\PsiBDC,\assign)$ as the
\textit{storage design}. The assignment matrix $\bm P$ must satisfy
the following conditions.
\begin{enumerate}
\item The entries of each row of $\assign$ must sum up to the batch size, i.e.,
  \begin{equation}
    \notag
    \sum_{j=1}^T p_{i, j} = \frac{r}{\binom{K}{\eta q}},\; 1 \leq i \leq \binom{K}{\eta q}.
  \end{equation}

\item The entries of each column of $\assign$ must sum up to the number of rows per
  partition, i.e.,
  \begin{equation}
    \notag
    \sum_{i=1}^{\binom{K}{\eta q}} p_{i, j} = \frac{r}{T},\; 1 \leq j \leq T.
  \end{equation}
\end{enumerate}

\begin{figure}[t!]
  \resizebox{1\columnwidth}{!}{
    \includegraphics{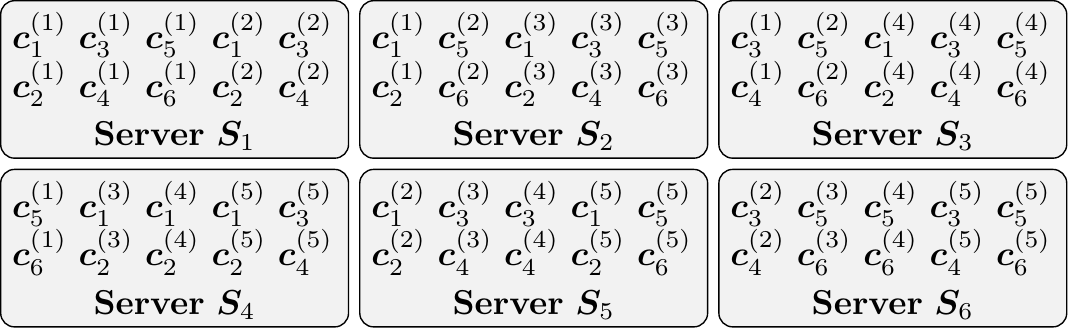}
  }
  \vspace{-2ex}
  \caption{Storage design for $m=20$, $N=4$, $K=6$, $q=4$, $\eta=1/2$, and
    $T=5$.}
  \label{fig:example_storage}
  \vspace{-2ex}
\end{figure}

\begin{figure}[t!]
  \resizebox{1\columnwidth}{!}{
    \includegraphics{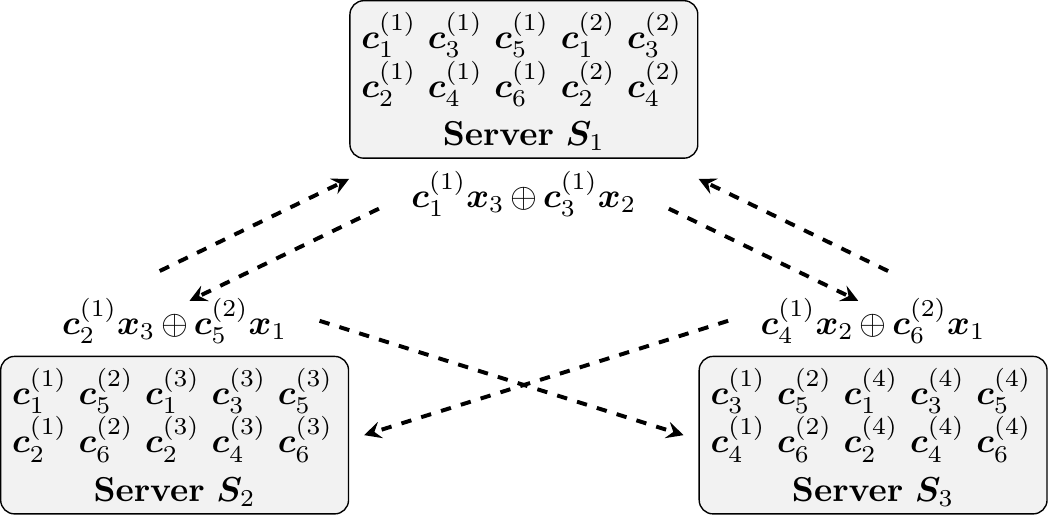}
  }
  \vspace{-2ex}
  \caption{Multicasting coded values between servers $S_1$, $S_2$, and $S_3$.}
  \label{fig:example_shuffling}
  \vspace{-2ex}
\end{figure}

We clarify the assignment of coded rows to batches and the coded
computing scheme in the following example.
\begin{example}[$m=20$, $N=4$, $K=6$, $q=4$, $\eta=1/2$, $T=5$]
  \label{ex:Example}

  For these parameters, there are $r/T=6$ coded rows per partition, of
  which $m/T=4$ are sufficient for decoding, and $\binom{K}{\eta q}=15$
  batches, each containing $r/ \binom{K}{\eta q}=2$ coded rows. We
  construct the storage design shown in Fig.~\ref{fig:example_storage}
  with $\binom{K}{\eta q} \times T = 15 \times 5$ assignment matrix
  \begin{equation}\label{eq:storage_design}
    \assign = \bordermatrix{
      ~ & 1 & 2 & 3 & 4 & 5 \cr
      (S_1, S_2) & 2 & 0 & 0 & 0 & 0 \cr
      (S_1, S_3) & 2 & 0 & 0 & 0 & 0 \cr
      (S_1, S_4) & 2 & 0 & 0 & 0 & 0 \cr
      (S_1, S_5) & 0 & 2 & 0 & 0 & 0 \cr
      ~~~~\vdots & & & \vdots & & \cr
      (S_4, S_6) & 0 & 0 & 0 & 0 & 2 \cr
      (S_5, S_6) & 0 & 0 & 0 & 0 & 2 \cr},
  \end{equation}%
  \noindent
  where rows are labeled by the subset of servers the batch is stored
  at, and columns are labeled by the partition index. In this case
  rows $\c_1^{(1)}$ and $\c_2^{(1)}$ are assigned to batch $1$,
  $\c_3^{(1)}$ and $\c_4^{(1)}$ are assigned to batch $2$, and so
  on. For this storage design, any $g=4$ servers collectively store at
  least $4$ coded rows from each partition. However, some servers
  store more rows than needed to decode some partitions, suggesting
  that this storage design is suboptimal.

  Assume that $\G=\{S_1, S_2, S_3, S_4\}$ is the set of $g=4$ servers
  that finish their map computations first. Also, assign vector $\y_i$
  to server $S_i$, $i=1,2,3,4$. We illustrate the coded shuffling
  scheme for $\mathcal{S} = \{S_1, S_2, S_3\}$ in
  Fig.~\ref{fig:example_shuffling}. Server $S_1$ multicasts
  $\bm{c}_1^{(1)} \x_3 \mathop{\oplus} \bm{c}_3^{(1)} \x_2$ to $S_2$
  and $S_3$.  Since $S_2$ and $S_3$ can cancel $\bm{c}_1^{(1)} \x_3$
  and $\bm{c}_3^{(1)} \x_2$, respectively, both servers receive one
  needed intermediate value.  Similarly, $S_2$ multicasts
  $\bm{c}_2^{(1)} \x_3 \mathop{\oplus} \bm{c}_5^{(2)} \x_1$, while
  $S_3$ multicasts
  $\bm{c}_4^{(1)} \x_2 \mathop{\oplus} \bm{c}_6^{(2)} \x_1$. This
  process is repeated for $\mathcal{S} = \{S_2, S_3, S_4\}$,
  $\mathcal{S} = \{S_1, S_3, S_4\}$, and
  $\mathcal{S} = \{S_1, S_2, S_4\}$. After the shuffle phase, we have
  sent $12$ multicast messages and $30$ unicast messages, resulting in
  a communication load of $(12 + 30)/20/4=0.525$, a $50\%$ increase
  from the load of the unpartitioned scheme ($0.35$, given by
  \eqref{eq:unpartitioned_load}). In this case, $S_1$ received
  additional intermediate values from partition $2$, despite already
  storing enough, further indicating that the assignment in
  \eqref{eq:storage_design} is suboptimal.

\end{example}

\section{Performance of the Block-Diagonal Coding}
\label{sec:performance}


In this section, we analyze the impact of partitioning on the
performance. We also prove that we can partition up to the batch size,
i.e., $T = r/\binom{K}{\eta q}$, without increasing the communication
load and the computational delay of the map phase with respect to the
original scheme in \cite{Li2016}.

\subsection{Communication Load}
\label{sec:load}

For the unpartitioned scheme of \cite{Li2016}, $\G = \Q$, and the
number of remaining values that need to be unicasted after the
multicast phase is constant regardless which subset $\Q$ of servers
finish first their map computations.  However, for the BDC
(partitioned) scheme, both $g$ and the number of remaining unicasts
may vary.

For a given assignment matrix $\assign$ and a specific $\Q$, we denote
by $U_\Q^{(S)}(\assign)$ the number of remaining values needed after
the multicast phase by server $S \in \Q$, and by
\begin{equation}
  \label{eq:uq}
  U_\Q(\assign) \triangleq \sum_{S \in \Q} U_\Q^{(S)}(\assign)
\end{equation}
the total number of remaining values needed by the servers in
$\Q$. Note that both $U_\Q^{(S)}(\assign)$ and $U_\Q(\assign)$ depend
on the strategy used to finish the shuffle phase (see
Section~\ref{sec:shuffle_phase}). We remark that all sets $\Q$ are equally
likely. Let $\mathbb{Q}^q$ denote the superset of all sets
$\Q$. Furthermore, we denote by $L_{\mathbb{Q}} (\assign)$ the average
communication load of the messages that are unicasted after the
multicasting step (see Section~\ref{sec:shuffle_phase}), i.e.,
\begin{equation}
  \label{eq:lq}
  L_{\mathbb{Q}} (\assign) \triangleq
  \frac{1}{mN} \frac{1}{\left\vert \mathbb{Q}^q \right\vert}
  \sum_{\Q \in \mathbb{Q}^q} U_\Q(\assign).
\end{equation}
When needed we write $L^{(1)}_{\mathbb{Q}} (\assign)$ and
$L^{(2)}_{\mathbb{Q}} (\assign)$, where the superscript denotes the
strategy used to finish the shuffle phase. For a given storage design
$(\PsiBDC,\assign)$, the communication load of the BDC scheme is given
by \makeatletter \if@twocolumn
\begin{equation} \label{eq:objective_function_expected}
  \begin{split}
    & \fexp(\PsiBDC, \assign) = \\
    & \min\left( \sum_{j=s_q}^{\eta q} \frac{\alpha_j}{\phi(j)} +
      L^{(1)}_{\mathbb{Q}} (\assign),
      \sum_{j=s_q-1}^{\eta q} \frac{\alpha_j}{\phi(j)}
      + L^{(2)}_{\mathbb{Q}} (\assign) \right).
  \end{split}
\end{equation}
\else
\begin{equation} \label{eq:objective_function_expected}
  \fexp(\PsiBDC, \assign) =
  \min\left( \sum_{j=s_q}^{\eta q} \frac{\alpha_j}{\phi(j)} +
    L^{(1)}_{\mathbb{Q}} (\assign),
    \sum_{j=s_q-1}^{\eta q} \frac{\alpha_j}{\phi(j)}
    + L^{(2)}_{\mathbb{Q}} (\assign) \right).
\end{equation}
\fi
\makeatother

Note that the load due to the multicast phase is independent of the
level of partitioning. Furthermore, for the unpartitioned scheme
$L^{(2)}_{\mathbb{Q}} = 0$ by design.

We first explain how $U_\Q^{(S)}$ is evaluated. Let $\bm{u}_\Q^{(S)}$
be a vector of length $T$, where the $t$-th element is the number of
intermediate values from partition $t$ stored by server $S$ at the end
of the multicast phase. Furthermore, each row of $\assign$ corresponds
to a batch, and coded multicasting is made possible by storing each
batch at multiple servers. The intermediate values transmitted during
the multicast phase thus correspond to rows of $\assign$. {The vector
  $\bm{u}_\Q^{(S)}$ is then computed by adding some set of rows of
  $\assign$. The indices of the rows to add depend on $\Q$ and $S$
  (see Section~\ref{sec:shuffle_phase}).

We denote by $\left( \bm{u}_\Q^{(S)} \right)_t$ the $t$-th element of
the vector $\bm{u}_\Q^{(S)}$. The number of values $U_\Q^{(S)}$ is
given by adding the number of intermediate values still needed for
each partition, i.e.,
\begin{equation}
  \label{eq:uqs}
  U_\Q^{(S)} = \sum_{t=1}^T \max\left( \frac{m}{T} -
    \left( \bm{u}_\Q^{(S)} \right)_t, 0\right).
\end{equation}
Its sum over all $S \in \Q$ gives $U_\Q(\assign)$ (see
\eqref{eq:uq}). Averaging $U_\Q(\assign)$ over all $\Q$ and normalizing
yields $L_{\mathbb{Q}} (\assign)$ (see \eqref{eq:lq}).

\begin{example}[Computing $\bm{u}_\Q^{(S)}$]
  We consider the same system as in Example~\ref{ex:Example}. We again assume that
  $\G=\Q=\{S_1, S_2, S_3, S_4\}$ is the set of $g=q=4$ servers that finish their
  map computations first. During the multicast phase server $S_1$ receives the
  intermediate values in $\V_{\mathcal{S}\setminus S_1}^{(S_1)}$ for all sets
  $\mathcal{S}$ of cardinality $j + 1 = 3$ (see Section~\ref{sec:shuffle_phase}). In this case, we
  perform coded multicasting within the sets
  \begin{itemize}
  \item $\mathcal{S} = \{S_1, S_2, S_3\}$,
    $\V_{\mathcal{S} \setminus S_1}^{(S_1)} = \{\c_5^{(2)}\x_1, \c_6^{(2)}\x_1 \}$,
  \item $\mathcal{S} = \{S_1, S_2, S_4\}$,
    $\V_{\mathcal{S} \setminus S_1}^{(S_1)} = \{\c_1^{(3)}\x_1, \c_2^{(3)}\x_1 \}$,
  \item $\mathcal{S} = \{S_1, S_3, S_4\}$,
    $\V_{\mathcal{S} \setminus S_1}^{(S_1)} = \{\c_1^{(4)}\x_1, \c_2^{(4)}\x_1 \}$.
  \end{itemize}

  Note that $\V_{\{S_2, S_3\}}^{(S_1)}$ contains the intermediate
  values computed from the coded rows stored in the batch that labels
  the $6$-th row of the assignment matrix $\assign$. In the same
  manner, $\V_{\{S_2, S_4\}}^{(S_1)}$ and $\V_{\{S_3, S_4\}}^{(S_1)}$
  correspond to rows $7$ and $10$ of $\assign$,
  respectively. Furthermore, prior to the shuffle phase server $S_1$
  stores the batches corresponding to rows $1$ to $5$ of
  $\assign$. Thus, $\bm{u}_{\{S_1, S_2, S_3, S_4\}}^{(S_1)}$ is equal
  to the sum of rows $1, 2, 3, 4, 5, 6, 7$, and $10$ of $\assign$. In
  this case,
  $\bm{u}_{\{S_1, S_2, S_3, S_4\}}^{(S_1)} = (6, 6, 2, 2, 0)$, and
  $S_1$ needs $8$ more intermediate values, i.e.,
  $U_{\{S_1, S_2, S_3, S_4\}}^{(S_1)} = 8$. Computing
  $\bm{u}_\Q^{(S)}$ for arbitrary $\Q$ and $S$ then corresponds to
  summing the rows of $\assign$ corresponding to batches either stored
  by server $S$ prior to the shuffle phase or received by $S$ in the
  multicast phase. The row indices are computed as explained in
  Section~\ref{sec:shuffle_phase}.
\end{example}

For a given $\PsiBDC$, the assignment of rows into batches can be
formulated as an optimization problem, where one would like to
minimize $\fexp(\PsiBDC, \assign)$ over all assignments
$\assign$. More precisely, the optimization problem is
\begin{equation}
  \notag
  \min_{\assign \in \mathbb{P}} \fexp(\PsiBDC,\assign),
\end{equation}
where $\mathbb{P}$ is the set of all assignments $\assign$. This is a
computationally complex problem, since both the complexity of
evaluating the performance of a given assignment and the number of
assignments scale exponentially in the problem size (there are
$q \binom{K}{q}$ vectors $\bm{u}_\Q^{(S)}$). We address the
optimization of the assignment matrix $\assign$ in Section~\ref{sec:solvers}.

\subsection{Computational Delay}
\label{computational_delay}

We consider the delay incurred by the encoding, map, and reduce phases
(see Definition~\ref{def:D}). As in \cite{Li2016}, we do not consider the delay
incurred by the shuffle phase as the computations it requires are
simple in comparison. Note that in \cite{Li2016} only $\tmap$ is
considered, i.e., $D = \tmap$. However, one should not neglect the
computational delay incurred by the encoding and reduce phases. Thus,
we consider the overall computational delay
% $D = \tenc  + \tmap + \tred$.
\begin{equation}
  \notag
  D = \tenc  + \tmap + \tred.
\end{equation}
The encoding delay $\tenc$ is a function of the number of nonzero
elements of $\PsiBDC$. As there are at most $\frac{m}{T}$ nonzero
elements in each row of a block-diagonal encoding matrix, for an
encoding scheme with $T$ partitions we have
\begin{equation}
  \label{eq:encoding_complexity}
  \cencbdc \leq \frac{m}{T} rn \cm + \left( \frac{m}{T} - 1\right)
  rn \ca.
\end{equation}

The reduce phase consists of decoding the $N$ output vectors and hence
the delay it incurs depends on the underlying code and decoding
algorithm. We assume that each partition is encoded using a
Reed-Solomon (RS) code and is decoded using either the
Berlekamp-Massey (BM) algorithm or the FFT-based algorithm proposed in
\cite{Lin2016}, whichever yields the lowest complexity. To the best of
our knowledge the algorithm proposed in \cite{Lin2016} is the lowest
complexity algorithm for decoding long RS codes. We measure the
decoding complexity by its associated shifted-exponential parameter
$\sigma$ (see Section~\ref{sec:ProbRuntime}).

The number of field additions and multiplications required to decode
an $(r/T, m/T$) RS code using the BM algorithm is
$(r/T)\left(\xi (r/T) - 1\right)$ and $(r / T)^2 \xi$, respectively,
where $\xi$ is the fraction of erased symbols \cite{Garr2013}. With
$\xi$ upper bounded by $1 - \frac{q}{K}$ (the map phase terminates
when a fraction of at least $\frac{q}{K}$ symbols from each partition
is available), the complexity of decoding the $T$ partitions for all
$N$ output vectors is upper bounded as \makeatletter \if@twocolumn
\begin{equation}
  \label{eq:decoding_complexity}
  \credbdc^{\mathsf{BM}} \leq
  N \left(
    \ca \left( \frac{r^2 (1-\frac{q}{K})}{T} -r \right) +
    \cm \frac{r^2 (1 - \frac{q}{K})}{T}
  \right).
\end{equation}
\else
\begin{equation}
  \label{eq:decoding_complexity}
  \credbdc^{\mathsf{BM}} \leq
  N \left(
    \ca \left( \frac{r^2 (1-\frac{q}{K})}{T} -r \right) +
    \cm \frac{r^2 (1 - \frac{q}{K})}{T}
  \right).
\end{equation}
\fi
\makeatother

On the other hand, the FFT-based algorithm has complexity
$\mathcal{O}(r \log r)$ \cite{Lin2016}. We estimate the number of
additions and multiplications required for a given code length $r$ by
fitting a curve of the form $a+br\log_2(cr)$, where $(a, b, c)$ are
coefficients, to empiric results derived from the authors'
implementation of the algorithm. For additions the resulting
parameters are $(2, 8.5, 0.867)$ and for multiplications they are
$(2, 1, 4)$. The resulting curves diverge negligibly at the measured
points. The total decoding complexity for the FFT-based algorithm is
\makeatletter \if@twocolumn
\begin{equation}
  \label{eq:fft_decoding_complexity}
  \begin{split}
    \credbdc^{\mathsf{FFT}} = \;
    & NT \ca \left( 2 + \frac{8.5r}{T} \log_2\left(0.867r/T\right)
    \right)\\
    & + NT \cm \left( 2 + \frac{r}{T} \log_2\left(4r/T\right) \right).
  \end{split}
\end{equation}
\else%
\begin{equation}
  \label{eq:fft_decoding_complexity}
  \credbdc^{\mathsf{FFT}} = \;
  NT \ca \left( 2 + \frac{8.5r}{T} \log_2\left(0.867r/T\right) \right)\\
  + NT \cm \left( 2 + \frac{r}{T} \log_2\left(4r/T\right) \right).
\end{equation}
\fi%
\makeatother%
The encoding and decoding complexity of the unified scheme in
\cite{Li2016} is given by evaluating \eqref{eq:encoding_complexity} and
either \eqref{eq:decoding_complexity} or
\eqref{eq:fft_decoding_complexity} (whichever gives the lowest
complexity), respectively, for $T=1$. For the BDC scheme, by choosing
$T$ close to $r$ we can thus significantly lower the delay of the
encoding and reduce phases. On the other hand, the scheme in
\cite{Lee2017} uses codes of length proportional to the number of
servers $K$. The encoding and decoding complexity of the SC scheme in
\cite{Lee2017} is thus given by evaluating
\eqref{eq:encoding_complexity} and either \eqref{eq:decoding_complexity}
or \eqref{eq:fft_decoding_complexity} for $T=\frac{m}{q}$.

\subsection{Lossless Partitioning}
\label{sec:lossless}

\begin{theorem}
  \label{pr:performance}
  For $T \leq r/\binom{K}{\eta q}$, there
  exists an assignment matrix $\assign$ such that the communication load and the
  computational delay of the map phase are equal to those of the unpartitioned
  scheme.
\end{theorem}
\begin{IEEEproof}
  \label{pro:delay}
  The computational delay of the map phase is equal to that of the
  unpartitioned scheme if any $q$ servers hold enough coded rows to
  decode all partitions. For $T = r/\binom{K}{\eta q}$ we let $\assign$
  be a $\binom{K}{\eta q} \times T$ all-ones matrix and show that it
  has this property by repeating the argument from
  \cite[Sec. IV.B]{Li2016} for each partition. In this case, any set
  of $q$ servers collectively store $\frac{\eta q m}{T}$ rows from each
  partition, and since each coded row is stored by at most $\eta q$
  servers, any $q$ servers collectively store at least
  $\frac{\eta q m}{\eta q T} = \frac{m}{T}$ unique coded rows from each
  partition. The computational delay of the map phase is thus
  unchanged from the unpartitioned scheme. The communication load is
  unchanged if $U_\Q^{(S)}$ is equal to that of the unpartitioned
  scheme for all $\Q$ and $S$. The number of values needed
  $U_\Q^{(S)}$ is computed from $\bm{u}_\Q^{(S)}$ (see
  \eqref{eq:uqs}), which is the sum of $l$ rows of $\assign$, for some
  integer $l$. For the all-ones assignment matrix, because all rows of
  $\assign$ are identical, we have
  \begin{equation}
    \notag
    U_\Q^{(S)} = T \max \left( \frac{m}{T} - l, 0 \right) =
    \max \left( m - Tl, 0 \right),
  \end{equation}
  which is the number of remaining values for the unpartitioned
  scheme.

  Next, we consider the case where $T < r / \binom{K}{\eta q}$. First,
  consider the case $T = r / \binom{K}{\eta q} - j$, for some integer
  $j$, $0 \leq j < \frac{r}{2 \binom{K}{\eta q}}$. We first set all
  entries of $\assign$ equal to $1$. At this point, the total number
  of unique rows of $\C$ per partition stored by any set of $q$
  servers is at least
  \begin{equation}
    \label{eq:assign1}
    \frac{m}{r / \binom{K}{\eta q}} =
    \frac{m}{r / \binom{K}{\eta q} - j}
    \frac{r / \binom{K}{\eta q} - j}{r / \binom{K}{\eta q}} =
    \frac{m}{T} \frac{r / \binom{K}{\eta q} - j}{r / \binom{K}{\eta q}}.
  \end{equation}
  The number of coded rows per partition that are not yet assigned is given by
  $r/T$ multiplied by the fraction of partitions removed $\frac{j}{r /
    \binom{K}{\eta q}}$, i.e.,
  \begin{equation} \label{eq:assign2}
    \frac{1}{T} \frac{r j}{r / \binom{K}{\eta q}} =
    \frac{1}{T} \frac{ m \frac{K}{q} j}{r / \binom{K}{\eta q}}.
  \end{equation}
  We assign these rows to batches such that an equal number of coded
  rows is assigned to each of the $K$ servers, which is always
  possible due to the limitations imposed by the system model. Any set
  of $q$ servers will thus store a fraction $q/K$ of these rows. The
  total number of unique coded rows per partition stored among any set
  of $q$ servers is then lower bounded by the sum of
  \eqref{eq:assign2} weighted by $q/K$ and \eqref{eq:assign1}, i.e.,
  \begin{equation}
    \notag
    \frac{m}{T} \left(
      \frac{r / \binom{K}{\eta q} - j}{r / \binom{K}{\eta q}} +
      \frac{\frac{K}{q}j}{r / \binom{K}{\eta q}} \frac{q}{K}
    \right) =
    \frac{m}{T},
  \end{equation}
  showing that it is possible to decode all partitions using the coded rows
  stored over any set of $q$ servers.

  The communication load is unchanged with respect to the case where
  the number of partitions is $r / \binom{K}{\eta q}$ if and only if no
  server receives rows it does not need in the multicast phase. Due to
  decreasing the number of partitions from $r / \binom{K}{\eta q}$ to
  $T = r / \binom{K}{\eta q} - j$, we increase the number of coded rows
  needed to decode each partition by
  \begin{equation}
    \label{eq:assign3}
    \frac{m}{T} - \frac{m}{r / \binom{K}{\eta q}} =
    \frac{1}{T}  \frac{m j}{r / \binom{K}{\eta q}}.
  \end{equation}
  Furthermore, reducing the number of partitions increases the number
  of coded rows per partition stored among any set of $q$ servers (see
  \eqref{eq:assign2} and the following text) by
  \begin{equation}
    \label{eq:assign4}
    \frac{1}{T} \frac{m j}{r / \binom{K}{\eta q}}.
  \end{equation}
  Note that the number of additional rows needed to decode each
  partition (see \eqref{eq:assign3}) is greater than or equal to the
  number of additional rows stored among the $q$ servers (see
  \eqref{eq:assign4}). It is thus impossible that too many coded rows
  are delivered for any partition.

  Second, we consider the case
  $T = \frac{r / \binom{K}{\eta q}-j}{i}$, where $j$ is chosen as for
  the first case above and where $i$ is a positive integer. Now, we
  first set all elements of $\assign$ to $i$.  At this point the
  number of unique rows of $\C$ per partition stored by any set of $q$
  servers is given by \eqref{eq:assign1} multiplied by a factor $i$
  (since we set each element of $\assign$ to $i$ instead of
  one). Furthermore, the number of coded rows per partition that are
  not yet assigned is given by \eqref{eq:assign2}. Therefore, by using
  the same strategy as for $i=1$ and assigning the remaining rows to
  batches such that an equal number of rows is assigned to each of the
  $K$ servers, we are guaranteed that the communication load and the
  computational delay are unchanged also in this case.
\end{IEEEproof}

\section{Assignment Solvers}
\label{sec:solvers}

For $T \leq r/\binom{K}{\eta q}$ partitions, we can choose the
assignment matrix $\assign$ as described in the proof of
Theorem~\ref{pr:performance}. For the case where
$T > r/\binom{K}{\eta q}$, we propose two solvers for the problem of
assigning rows into batches: a heuristic solver that is fast even for
large problem instances, and a hybrid solver combining the heuristic
solver with a branch-and-bound solver. The branch-and-bound solver
produces an optimal assignment but is significantly slower, hence it
can be used as stand-alone only for small problem instances. We use a
dynamic programming approach to speed up the branch-and-bound solver
by caching $\bm{u}_\Q^{(S)}$ for all $S$ and $\Q \in \mathbb{Q}^q$. We
index each cached $\bm{u}_\Q^{(S)}$ by the batches it is computed
from. Whenever $U_\Q^{(S)}$ drops to $0$ due to assigning a row to a
batch, we remove the corresponding $\bm{u}_\Q^{(S)}$ from the
index. We also store a vector of length $T$ with the $i$-th entry
giving the number of vectors $\bm{u}_\Q^{(S)}$ that miss intermediate
values from the $i$-th partition. Specifically, the $i$-th element of
this vector is the number of vectors $\bm{u}_\Q^{(S)}$ for which the
$i$-th element is less than $\frac{m}{T}$. This allows us to
efficiently assess the impact on $L_{\mathbb{Q}} (\assign)$ due to
assigning a row to some batch. Since $\bm{u}_\Q^{(S)}$ is of length
$T$ and because the cardinality of $\Q$ and $\mathbb{Q}^q$ is $q$ and
$\binom{K}{q}$, respectively, the memory required to keep this index
scales as $\mathcal{O}\left(Tq\binom{K}{q}\right)$ and is thus only an
option for small problem instances.

For all solvers, we first label the batches lexiographically and then
optimize $\fexp$ in \eqref{eq:objective_function_expected}. For
example, for $\eta q=2$, we label the first batch by $S_1, S_2$, the
second by $S_1, S_3$, and so on. The solvers are available under the
Apache 2.0 license \cite{Severinson2017code}. We remark that choosing
$\assign$ is similar to the problem of designing the coded matrices
stored by each server in \cite{Qian2017}.

\subsection{Heuristic Solver}
The heuristic solver is inspired by the assignment matrices created by
the branch-and-bound solver for small instances. It creates an
assignment matrix $\assign$ in two steps. We first set each entry of
$\assign$ to
\begin{equation} \notag
  Y \triangleq \left\lfloor \frac{r}{\binom{K}{\eta q}\cdot T}\right\rfloor,
\end{equation}
thus assigning the first $\binom{K}{\eta q} Y$ rows of each partition to batches
such that each batch is assigned $Y T$ rows. Let $d=r/\binom{K}{\eta q}-Y T$ be
the number of rows that still need to be assigned to each batch. The
$r/T-\binom{K}{\eta q}Y$ rows per partition not assigned yet are assigned in the
second step as shown in Algorithm 1.
\begin{algorithm}[!t]
  \SetKwInOut{Input}{Input}
  \SetKwInOut{Output}{Output}
  \Input{$\bm P$, $d$, $K$, $T$, and $\eta q$}
  \For{$0\le a < d\binom{K}{\eta q}$}{\label{algcPoP:outerWhile}
    $i \leftarrow \lfloor a/d \rfloor +1$\\
    $j \leftarrow (a \bmod T)+1$\\
    $p_{i,j} \leftarrow p_{i,j}+1$
  }
  \KwRet{$\bm P$}
  \caption{Heuristic Assignment}
\end{algorithm}

Interestingly, for $T \leq r/\binom{K}{\eta q}$ the heuristic solver
creates an assignment matrix satisfying the requirements outlined in
the proof of Theorem~\ref{pr:performance}. In the special case of
$T = r/\binom{K}{\eta q}$, the all-ones matrix is produced.

\subsection{Branch-and-Bound Solver}
The branch-and-bound solver finds an optimal solution by recursively branching
at each batch for which there is more than one possible assignment and
considering all options. The solver is initially given an empty assignment
matrix, i.e., an all-zeros $\binom{K}{\eta q} \times T$ matrix. For each branch,
we lower bound the value of the objective function of any assignment in that
branch and only investigate branches with possibly better assignments. The
branch-and-bound operations given below are repeated until there are no more
potentially better solutions to consider.

\subsubsection{Branch}
For the first row of $\assign$ with remaining assignments, branch on every
available assignment for that row. More precisely, find the smallest index $i$
of a row of the assignment matrix $\assign$ whose entries do not sum up to the batch size,
i.e.,
\begin{equation}
  \notag
  \sum_{j=1}^T p_{i, j} < \frac{r}{\binom{K}{\eta q}}.
\end{equation}
For row $i$, branch on incrementing the element $p_{i, j}$ by $1$ for all columns
(with index $j$) such that their entries do not sum up to the number of coded rows per partition, i.e.,
\begin{equation}
  \notag
  \sum_{i=1}^{\binom{K}{\eta q}} p_{i, j} < \frac{r}{T}.
\end{equation}

\subsubsection{Bound}
We use a dynamic programming approach to lower bound $\fexp$ for a
subtree.  Specifically, for each row $i$ and column $j$ of $\assign$,
we store the number of vectors $\bm{u}_\Q^{(S)}$ that are indexed by
row $i$ and where the $j$-th element satisfies
\begin{equation}
  \notag
  \left(\frac{m}{T} - \left(\bm{u}_\Q^{(S)}\right)_j\right) > 0.
\end{equation}

Assigning a coded row to a batch can at most reduce $\fexp$ by $1 / \left(m N
  \left\vert\mathbb{Q}^q\right\vert \right)$ for each $\bm{u}_\Q^{(S)}$ indexed
by that batch. We compute the bound by assuming that no $\bm{u}_\Q^{(S)}$ will
be removed from the index for any subsequent assignment.

\subsection{Hybrid Solver}
The branch-and-bound solver can only be used by itself for small instances.
However, it can be used to complete a \emph{partial} assignment matrix, i.e., a
matrix $\assign$ for which not all rows have entries that sum up to the batch size. The
branch-and-bound solver then completes the assignment optimally. We first find a
candidate solution using the heuristic solver and then iteratively improve it
using the branch-and-bound solver. In particular, we decrement by $1$ a random
set of entries of $\assign$ and then use the branch-and-bound solver to
reassign the corresponding rows optimally. We repeat this process until the
average improvement between iterations drops below some threshold.

\section{Luby Transform Codes}
\label{sec:rateless}

In this section, we consider LT codes \cite{Luby2002} for use in
distributed computing. Specifically, we consider a distributed
computing system where $\enc$ is an LT code encoding matrix, denoted
by $\PsiLT$, of fixed rate $\frac{m}{r}$.  As explained in
Section~\ref{sec:SystemModel}, we divide the $r$ coded rows of
$\C = \PsiLT \A$ into $\binom{K}{\eta q}$ disjoint batches, each of
which is stored at a unique subset of size $\eta
q$ of the $K$ servers. For
this scheme, due to the random nature of LT codes, we can assign coded
rows to batches randomly. The distributed computation is carried out
as explained in Section~\ref{sec:computing_model}, i.e., we wait for the
fastest $g \geq q$ servers to complete their respective computations
in the map phase, perform coded multicasting during the shuffle phase,
and carry out the decoding of the $N$ output vectors in the reduce
phase.

Let $\Omega$ denote the degree distribution and $\Omega(d)$ the
probability of degree $d$. Also, let $\bar{\Omega}$ be the average
degree. Then, each row of the encoding matrix $\PsiLT$ is constructed
in the following manner. Uniformly at random select $d$ unique entries
of the row, where $d$ is drawn from the distribution $\Omega$.  For
each of these $d$ entries, assign to it a nonzero element selected
uniformly at random from $\mathbb{F}_{2^l}$. Specifically, we consider
the case where $\Omega$ is the robust Soliton distribution
parameterized by $M$ and $\delta$, where $M$ is the location of the
spike of the robust component and $\delta$ is a parameter for tuning
the decoding failure probability for a given $M$ \cite{Luby2002}.

\subsection{Inactivation Decoding}
\label{sec:lt_decoding}

We assume that decoding is performed using inactivation decoding
\cite{rfc6330}. Inactivation decoding is an efficient maximum
likelihood decoding algorithm that combines iterative decoding with
optimal decoding in a two-step fashion and is widely used in practice. As suggested in
\cite{rfc6330}, we assume that the optimal decoding phase is
performed by Gaussian elimination. In particular, iterative decoding
is used until the ripple is empty, i.e., until there are no coded
symbols of degree $1$, at which point an input symbol is
inactivated. The iterative decoder is then restarted to produce a
solution in terms of the inactivated symbol. This procedure is
repeated until all input symbols are either decoded or
inactivated. Note that the value of some input symbols may be
expressed in terms of the values of the inactivated symbols at this point. Finally,
optimal decoding of the inactivated symbols is performed via Gaussian
elimination, and the decoded values are back-substituted into the
decoded input symbols that depend on them. The decoding
  schedule has a large performance impact. Our implementation follows
  the recommendations in \cite{rfc6330}. It is important to tune the
parameters $M$ and $\delta$ to minimize the number of inactivations.

Due to the nature of LT codes, we need to collect $m(1+\epsilon)$
intermediate values for each vector $\y$ before decoding. We refer to
$\epsilon$ as the overhead. Under inactivation decoding, and for a
given overhead $\epsilon$, the probability of decoding failure with an
overhead of at most $\epsilon$, denoted by $\Pf(\epsilon)$, is
lower bounded by \cite{Schotsch2013}
\begin{equation}
  \label{eq:lt_failure}
  \Pf(\epsilon) \geq \sum_{i=1}^m (-1)^{i+1} \binom{m}{i} \left(
    \sum_{d=1}^m \Omega(d) \frac{\binom{m-i}{d}}{\binom{m}{d}}
  \right)^{m(1+\epsilon)}.
\end{equation}
Note that $\Pf(\epsilon)$ is the CDF for the random variable
\enquote{decoding is not possible at a given overhead $\epsilon$.}
Furthermore, the lower bound (\ref{eq:lt_failure}) well approximates
the failure probability for an overhead slightly larger than
$\epsilon = 0$. Denote by $F_\mathsf{DS}(\epsilon)$ the probability of
decoding being possible at an overhead of at most $\epsilon$. It follows that
% $F_\mathsf{DS}(\epsilon) = 1-\Pf(\epsilon)$.
\begin{equation}
  \notag
  F_\mathsf{DS}(\epsilon) = 1-\Pf(\epsilon).
\end{equation}
We find the decoding success probability density function (PDF) by
numerically differentiating $F_\mathsf{DS}(\epsilon)$.

\subsection{Code Design}
We design LT codes for a minimum overhead $\epsilon_\mathsf{min}$, i.e., we
collect at least $m(1+\epsilon_\mathsf{min})$ coded symbols from the servers
before attempting to decode, and a target failure probability $\Pft =
\Pf(\epsilon_\mathsf{min})$. We remark that increasing $\epsilon_\mathsf{min}$
and $\Pft$ leads to a lower average degree $\bar{\Omega}$, and thus to less
complex encoding and decoding and subsequently to a lower computational delay
for encoding and decoding. The tradeoff is that the communication load increases
as more intermediate values need to be transferred over the network on average.
Furthermore, increasing $\epsilon_\mathsf{min}$ and $\Pft$ may increase the
average number of servers $g$ required to decode. We thus need to balance the
computational delay of the encoding and reduce phases against that of the map
phase to achieve a low overall computational delay. Furthermore, waiting for
more than $g=q$ servers typically increases the overall computational delay by
more than what is saved by the less complex encoding and decoding given by the
larger $\epsilon_\mathsf{min}$ and $\Pft$. We thus choose
$\epsilon_\mathsf{min}$ and $\Pft$ such that decoding is possible with high
probability using the number of coded rows stored at any set of $q$ servers.
Note that the overhead $\epsilon$ required for decoding may be larger than
$\epsilon_\mathsf{min}$. We take this into account by numerically integrating
the decoding success PDF multiplied by the performance of the scheme as a
function of the overhead $\epsilon$.

For a given $\epsilon_\mathsf{min}$ and $\Pft$, we find a pair
$(M,\delta)$ that minimizes the decoding complexity (see
Section~\ref{sec:lt_delay}) under the constraint that
$\Pf(\epsilon_\mathsf{min}) \approx \Pft$. Essentially, we minimize
the computational delay of the reduce phase for a fixed delay of the
map phase. We remark that LT codes with low decoding complexity have a
low average degree $\bar{\Omega}$, and thus also low encoding
complexity. Note that for a given $M$, decreasing $\delta$ lowers the
failure probability, but also increases the decoding complexity. We
find good pairs $(M, \delta)$ by selecting through binary search the
largest $M$ such that there exists a $\delta$ for which the lower
bound on $\Pf(\epsilon_\mathsf{min})$ in \eqref{eq:lt_failure} is
approximately equal to $\Pft$. This heuristic produces codes with
complexity very close to those found using basin-hopping
\cite{Wales1997} combined with the Powell optimization method
\cite{Powell1964}.

\subsection{Computational Delay}
\label{sec:lt_delay}

There are on average $\bar{\Omega}$ nonzero entries in each row of the LT code
encoding matrix. The LT code encoding complexity is thus given by
\begin{equation}
  \notag
  \cenclt = \bar{\Omega} rn \cm + (\bar{\Omega}-1) rn \ca.
\end{equation}
We simulate the complexity of the decoding
$\sigma_\mathsf{reduce, LT}$. Furthermore, we assume that the decoding
complexity depends only on $\epsilon_\mathsf{min}$, i.e., we evaluate
the decoding complexity only at $\epsilon = \epsilon_\mathsf{min}$,
and simulate the number of servers $g$ required for a given overhead
$\epsilon$.

\subsection{Communication Load}

The coded multicasting scheme (see Section~\ref{sec:shuffle_phase}) is
designed for the case where we need $m$ intermediate values per vector
$\y$. Here, we tune it for the case where we instead need at least
$m(1+\epsilon_\mathsf{min})$ intermediate values by increasing the
number of coded multicast messages sent.  Note that the coded
multicasting scheme is greedy in the sense that it starts by
multicasting coded messages to the largest possible number of
recipients and then gradually lowers the number of recipients.
Specifically, we perform the shuffle phase with (see \eqref{eq:sq})
\begin{equation}
  \notag
  s_{q, \mathsf{LT}} \triangleq \text{inf}\left(
    s : \sum_{j=s}^{\eta q} \alpha_j \leq (1+\epsilon_\mathsf{min}) - \eta
  \right).
\end{equation}
The communication load of the LT code-based scheme for a given
$\epsilon \geq \epsilon_{\mathsf{min}}$ is then given by
\makeatletter
\if@twocolumn
\begin{equation}
  \notag
  \begin{split}
    L_{\mathsf{LT}}
    & = \min \left(
      \sum_{j=s_{q, \mathsf{LT}}}^{\eta q}
      \frac{\alpha_j}{\phi(j)}
      + (1+\epsilon) - \eta - \sum_{j=s_{q, \mathsf{LT}}}^{\eta q}
      \alpha_j \right.,\\
    & \left.  \sum_{j=s_{q, \mathsf{LT}} -1 }^{\eta q}
      \frac{\alpha_j}{\phi(j)} + \max\left( (1+\epsilon) - \eta
        - \sum_{j=s_{q, \mathsf{LT}} - 1}^{\eta q} \alpha_j, 0 \right)
    \right).
  \end{split}
\end{equation}
\else
\begin{equation}
  \notag
  \begin{split}
    L_{\mathsf{LT}}
    & =
    \min \left(
      \sum_{j=s_{q, \mathsf{LT}}}^{\eta q}
      \frac{\alpha_j}{\phi(j)} +
      (1+\epsilon) - \eta - \sum_{j=s_{q, \mathsf{LT}}}^{\eta q}
      \alpha_j \right.,\\
    & \left.
      \sum_{j=s_{q, \mathsf{LT}} -1 }^{\eta q}
      \frac{\alpha_j}{\phi(j)} +
      \max\left(
        (1+\epsilon) - \eta - \sum_{j=s_{q, \mathsf{LT}} - 1}^{\eta q}
        \alpha_j,
        0 \right) \right).
  \end{split}
\end{equation}
\fi
\makeatother

\subsection{Partitioning of the LT Code-Based Scheme}

We can apply partitioning to the LT code-based scheme in the same manner as for the BDC scheme.
Specifically, we consider a block-diagonal encoding matrix $\PsiBDCLT$, where
the blocks $\bm{\psi}_1, \ldots, \bm{\psi}_{T}$ are LT code encoding
matrices. In particular, we consider the case where the number of partitions
$T$ is equal to the partitioning limit of Theorem~\ref{pr:performance},
i.e., $T = r/\binom{K}{\eta q}$. In this case the all-ones assignment
matrix $\assign$ introduced in the proof of Theorem~\ref{pr:performance} is a valid matrix. By
using this assignment matrix and identical encoding matrices for each
of the partitions, i.e., $\bm{\psi}_i=\bm{\psi}$, $i=1, \ldots, T$, the
encoding and decoding complexity of each partition is identical regardless of
which set of servers $\G$ first completes the map phase. Furthermore, by the
same argument as in the proof of Theorem~\ref{pr:performance}, we are guaranteed that if any
partition can be decoded using the coded rows stored at the set of servers $\G$,
all other partitions can also be decoded.

\section{Numerical Results}
\label{sec:results}

\begin{figure}[t!]
  \hspace{-2ex} \resizebox{1.05\columnwidth}{!}{
    \includegraphics{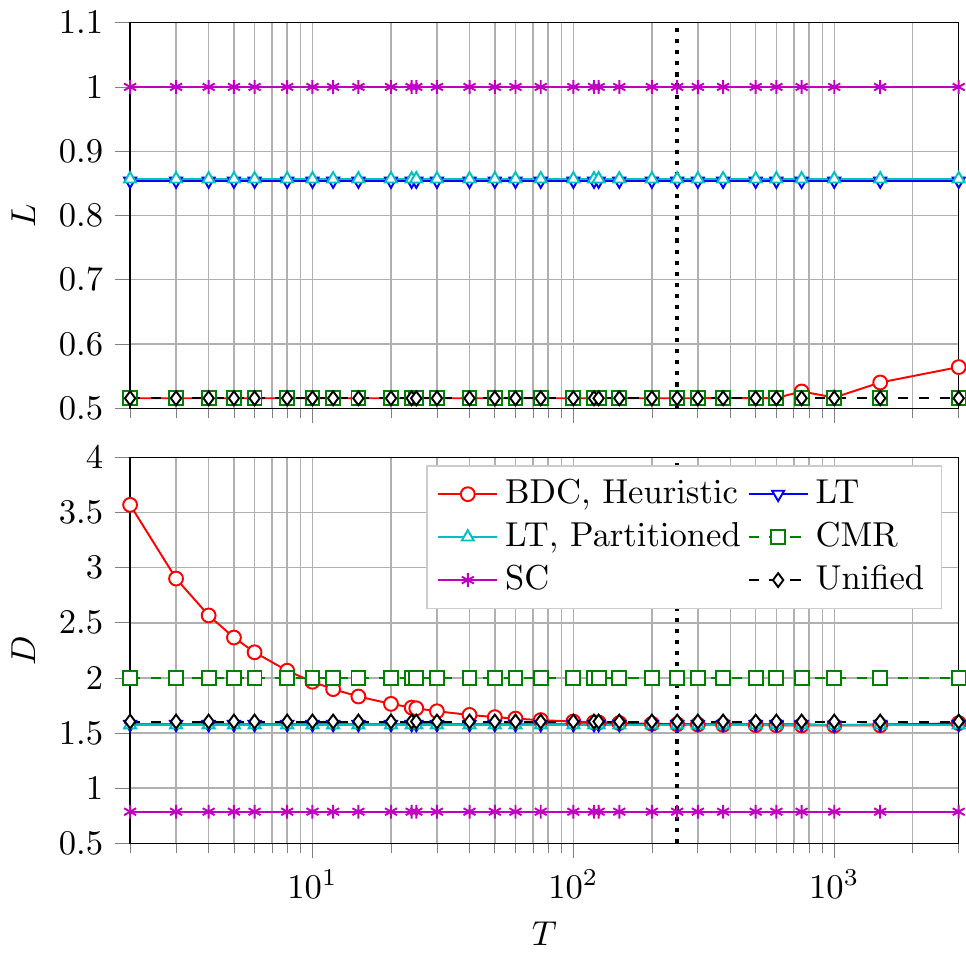}
  }
  \vspace{-5ex}
  \caption{The tradeoff between partitioning and performance for $m=6000$,
    $n=6000$, $K=9$, $q=6$, $N=6000$, and $\eta=1/3$.}
  \vspace{-3ex}
  \label{fig:numerical_partitions}
\end{figure}
\begin{figure}[t!]
  \hspace{-2ex} \resizebox{1.05\columnwidth}{!}{
    \includegraphics{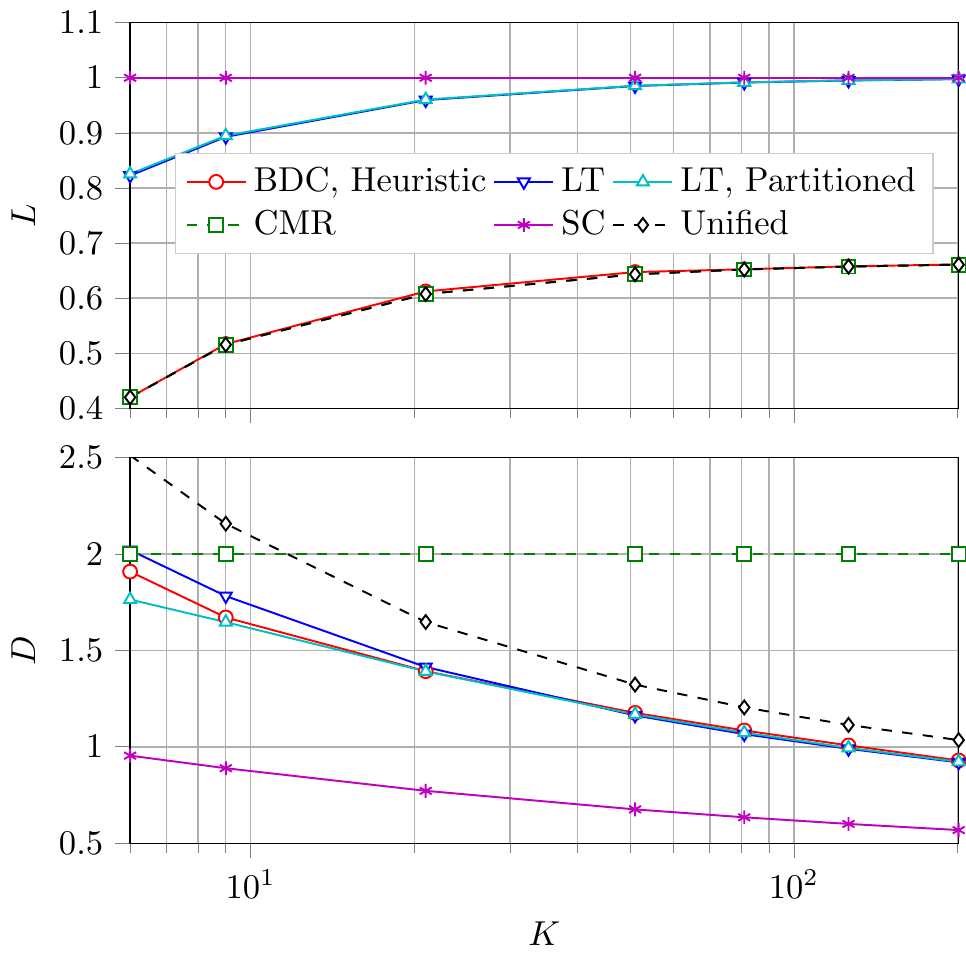}
  }
  \vspace{-5ex}
  \caption{Performance dependence on system size for $\eta q=2$,
    $n=m/100$, $\eta m=2000$, code rate $m/r=2/3$, and
    $N=500q$ vectors.}
  \vspace{-3ex}
  \label{fig:numerical_size}
\end{figure}%

We present numerical results for the proposed BDC and LT code-based
schemes and compare them with the schemes in
\cite{Li2015,Lee2017,Li2016}. Furthermore, we compare the performance
of the BDC scheme with assignment $\assign$ produced by the heuristic
and hybrid solvers. We also evaluate the performance of the LT
code-based scheme for different $\Pft$ and
$\epsilon_\mathsf{min}$. For each plot, the field size is equal to one
more than the largest number of coded rows considered for that plot,
$r+1$, rounded up to the closest power of $2$. The results, except
those in Fig.~\ref{fig:deadline}, are normalized by the performance of
the uncoded scheme. Unless stated otherwise, the assignment $\assign$
is given by the heuristic solver. As in \cite{Li2016}, we assume that
$\phi(j) = j$.

\subsection{Coded Computing Comparison}

In Fig.~\ref{fig:numerical_partitions}, we depict the communication load
$L$ (see Definition~\ref{def:L}) and the computational delay $D$ (see
Definition~\ref{def:D}) as a function of the number of partitions, $T$. The
system parameters are $m=6000$, $n=6000$, $K=9$, $q=6$, $N=6000$, and
$\eta=1/3$. The parameters of the CMR and SC schemes are
$q_{\mathsf{CMR}} = 9$, $\eta_{\mathsf{CMR}} = \frac{2}{9}$, and
$\eta_{\mathsf{SC}} = \frac{1}{6}$. The minimum overhead for the LT
code-based scheme is $\epsilon_\mathsf{min}=0.3$ and its target
failure probability is $\Pft=0.1$. For up to
$r / \binom{K}{\eta q}=250$ partitions (marked by the vertical dotted
line), the BDC scheme does not incur any loss in $\tmap$ and
communication load with respect to the unified scheme (see
Theorem~\ref{pr:performance}). Furthermore, the BDC scheme yields about a
$2$\% lower delay compared to the unified scheme for $T=1000$. The
delay of the LT code-based scheme is slightly worse than that of the
BDC scheme, and the load is about $65\%$ higher (for
$T=250$). Partitioning the LT code-based scheme increases the
communication load and reduces the computational delay by about
$0.5$\%. We remark that the number of partitions for the LT code-based
scheme is fixed at $ r/\binom{K}{\eta q}$. For heavy partitioning of
the BDC scheme, a tradeoff between partitioning level, communication
load, and map phase delay is observed.  For example, with $3000$
partitions (the maximum possible), there is about a $10\%$ increase in
communication load over the unified scheme.  Note that the gain in
computational delay saturates, thus there is no reason to partition
beyond a given level. The load of the SC scheme is about twice that of
our proposed schemes and the delay is about half. Finally, the delay
of the BDC and the LT code-based scheme is about $25\%$ lower compared
to the CMR scheme for $T > 100$.

In Fig.~\ref{fig:numerical_size}, we plot the performance for a constant
$\eta q = 2$, $n=m/100$, $\eta m = 2000$, code rate $m/r=2/3$, and
$N=500q$ vectors as a function of the number of servers, $K$. The
ratio $m/n$ is motivated by machine learning applications, where the
number of rows and columns often represent the number of samples and
features, respectively. Note that the number of arithmetic operations
performed by each server in the map phase increases with $K$. We
choose the number of partitions $T$ that minimizes the delay under the
constraint that the communication load is at most $1$\% higher
compared to the unified scheme. The parameters of the LT code-based
scheme are $\epsilon_\mathsf{min}=0.335$ and $\Pft=0.1$. The results
shown are averages over $1000$ randomly generated realizations of
$\G$.  Our proposed BDC scheme outperforms the unified scheme in terms
of computational delay by between about $25$\% (for $K=6$) and $10$\%
(for $K=201$). Furthermore, the delay of both the BDC and LT
code-based schemes are about $50$\% lower than that of the CMR scheme
for $K=201$. For $K=6$ the computational delay of the unpartitioned
and partitioned LT code-based schemes is about $5$\% higher and $8$\%
lower compared to the BDC scheme, respectively. For $K=201$ the delay
of the LT code-based scheme is about $1$\% lower than that of the BDC
scheme. However, the communication load is about $45\%$
higher. Finally, the communication load of the BDC scheme is between
about $42\%$ (for $K=6$) and $66\%$ (for $K=201$) of that of the SC
scheme.

\begin{figure}[t!]
  \hspace{-2ex} \resizebox{1.03\columnwidth}{!}{
    \includegraphics{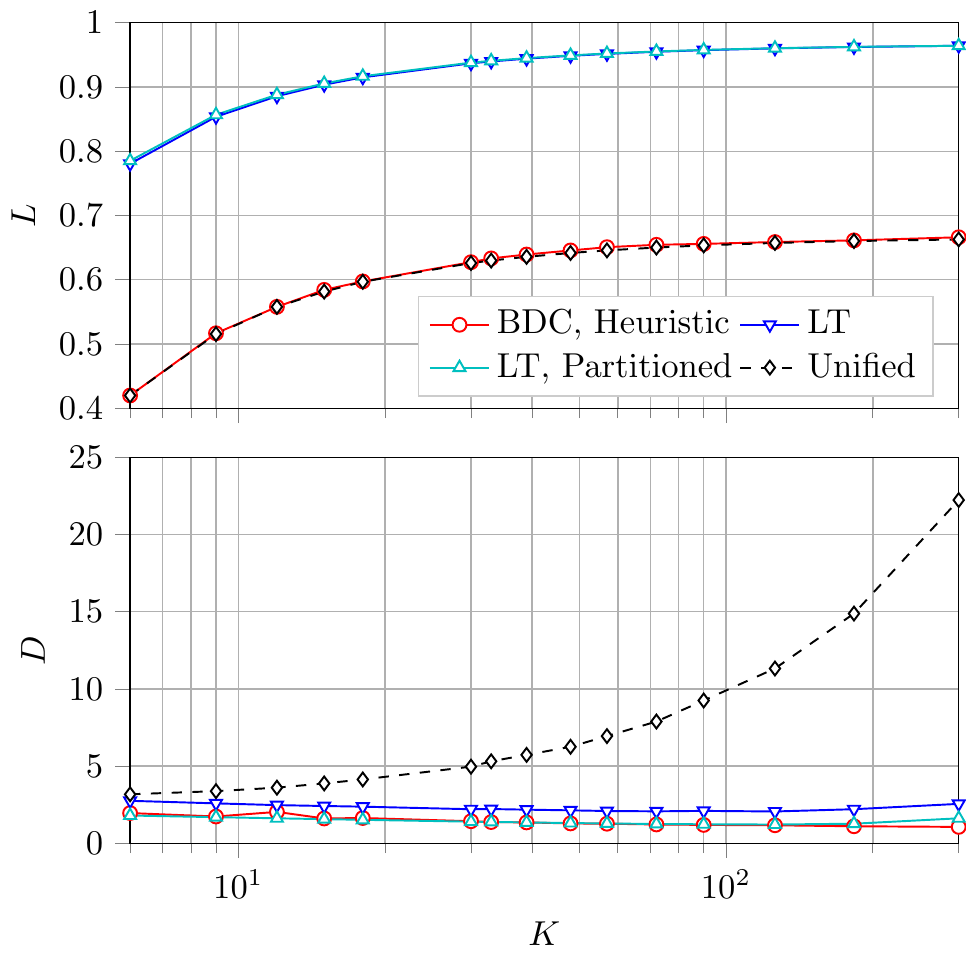}
  }
  \vspace{-5ex}
  \caption{Performance dependence on system size with constant
    complexity of the map phase per server, $m/r = 2/3$, $\eta q = 2$,
    $n=m/100$, and $N=n$.}
  \vspace{-3ex}
  \label{fig:workload}
\end{figure}
\begin{figure}[t!]
  \hspace{-2ex} \resizebox{1.05\columnwidth}{!}{
    \includegraphics{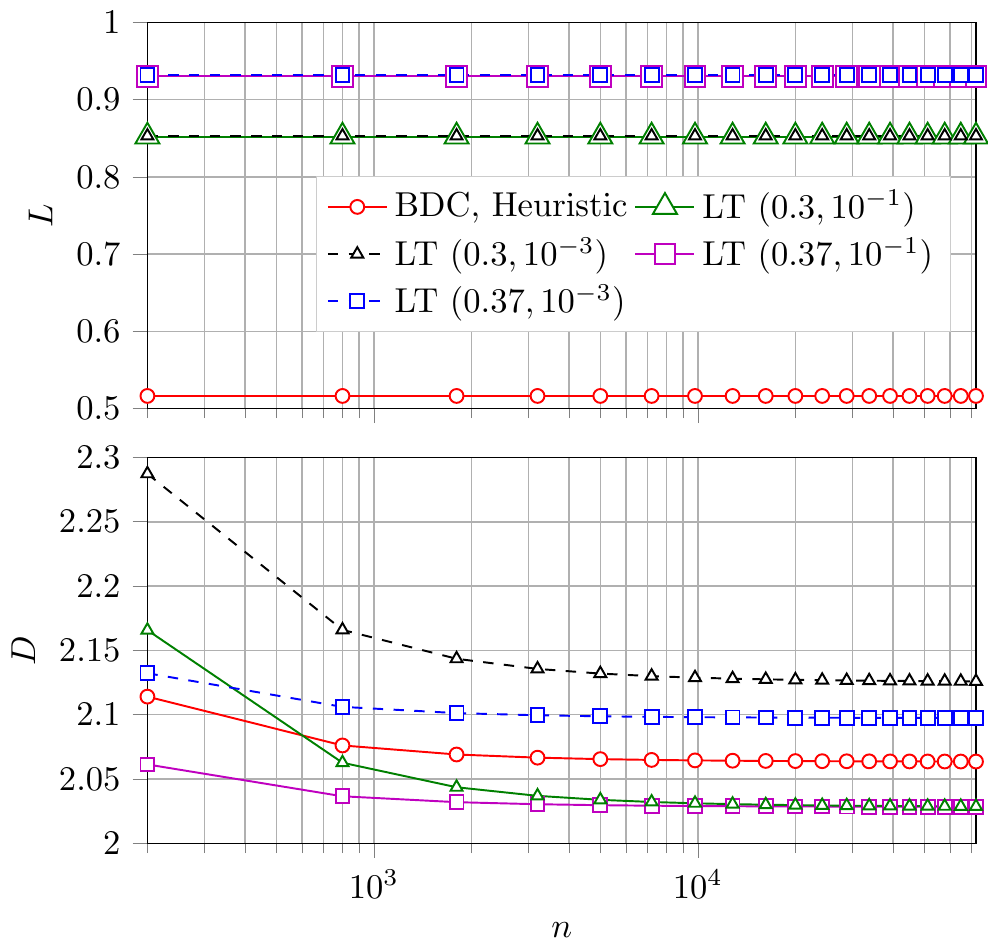}
  }
  \vspace{-4.2ex}
  \caption{Performance dependence on the number of columns $n$ of
    $\bm A$ for $m=2400$, $K=9$, $q=6$, $N=60$, $T=240$, and
    $\eta=1/3$. The parameters of the LT code-based scheme are given in
    the legend as $(\epsilon_\mathsf{min}, \Pft)$.}
  \vspace{-3ex}
  \label{fig:lt_ratio}
\end{figure}

In Fig.~\ref{fig:workload}, we show the performance for code rate
$m/r = 2/3$, $\eta q = 2$, and a fixed workload per server as a
function of $K$. Specifically, we fix the number of additions and
multiplications computed by each server in the map phase to $10^8$
($\pm 5$\% to find valid parameters) and scale $m, n, N$ with $K$.
The number of rows $m$ of $\bm A$ takes values between $12600$ and $59800$, and  we let $n=m/100$ and
$N=n$. The number of partitions $T$ is selected in the same way as for
Fig.~\ref{fig:numerical_size}.  The results shown are averages over $1000$
randomly generated realizations of $\G$. The computational delay of
the unified scheme is about a factor $20$ higher than that of the BDC
scheme for $K=300$. The computational delay of the partitioned LT
code-based scheme is similar to that of the BDC scheme, while the
delay of the unpartitioned LT code-based scheme is about $60$\%
higher. Furthermore, the communication load of the LT code-based
scheme is about $45$\% higher compared to those of the unified and BDC
schemes.

In Fig.~\ref{fig:lt_ratio}, we plot the performance of the BDC and LT
code-based schemes as a function of the number of columns $n$. The
system parameters are $m=2400$, $K=9$, $q=6$, $N=60$, $T=240$, and
$\eta=1/3$. The communication load of the LT code-based scheme depends
primarily on the minimum overhead $\epsilon_\mathsf{min}$ and the
computational delay primarily on the target failure probability
$\Pft$. We remark that a higher $\Pft$ allows for using codes with
lower average degree and thus less complex encoding and decoding. For
$n=20000$, the computational delay of the LT code-based scheme with
$\Pft=0.1$ is about $1.5$\% lower than that of the BDC scheme. For
$\Pft=0.001$, the computational delay is about $3$\% and $1.5$\%
higher than that of the BDC scheme when $\epsilon_\mathsf{min}=0.3$
and $\epsilon_\mathsf{min}=0.37$, respectively. On the other hand, the
communication load of the LT code-based scheme with
$\epsilon_\mathsf{min}=0.3$ and $\epsilon_\mathsf{min}=0.37$ is about
$41\%$ and $44\%$ higher than that of the BDC scheme, respectively.

\subsection{Assignment Solver Comparison}

\begin{figure}
  \hspace{-2ex} \resizebox{1.05\columnwidth}{!}{
    \includegraphics{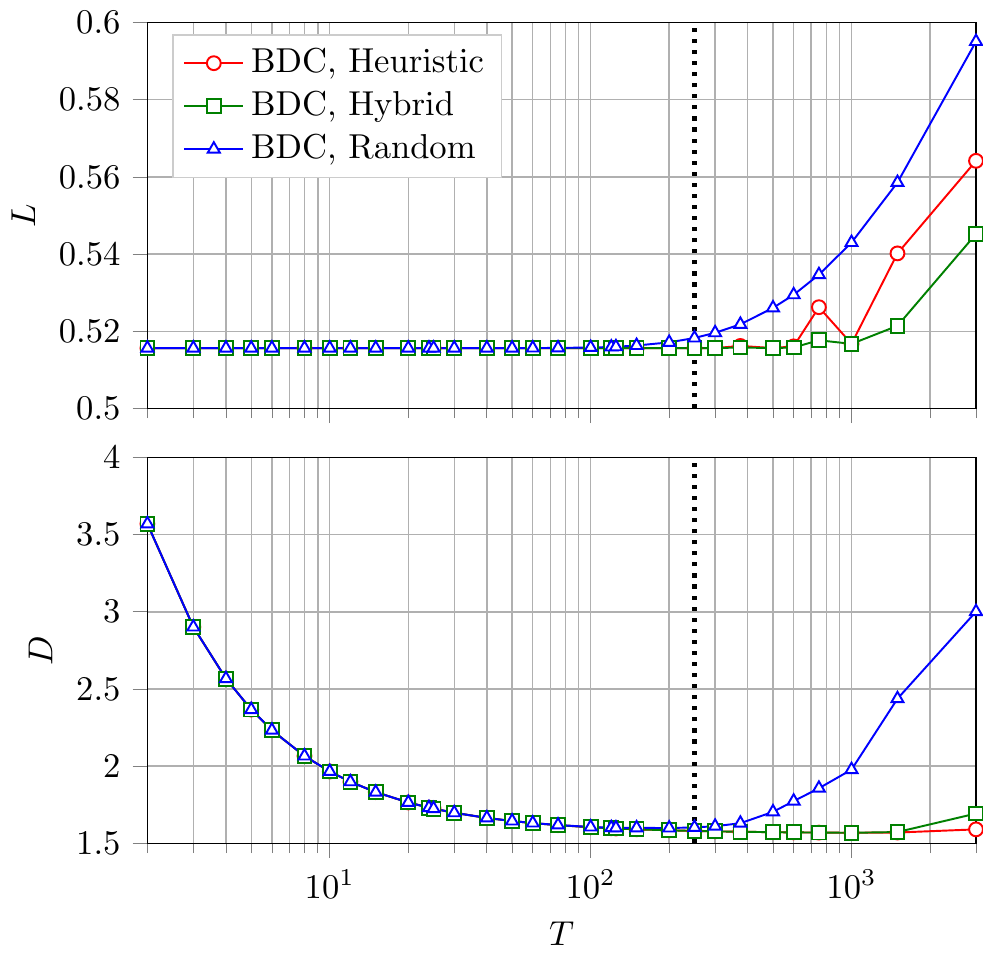}
  }
  \vspace{-4.5ex}
  \caption{Solver performance as a function of partitioning for
    $m=6000$, $n=6000$, $K=9$, $q=6$, $N=6000$, and $\eta=1/3$.}
  \vspace{-0.1ex}
  \label{fig:solvers_partitions}
\end{figure}
\begin{figure}
  \hspace{-2ex} \resizebox{1.05\columnwidth}{!}{
    \includegraphics{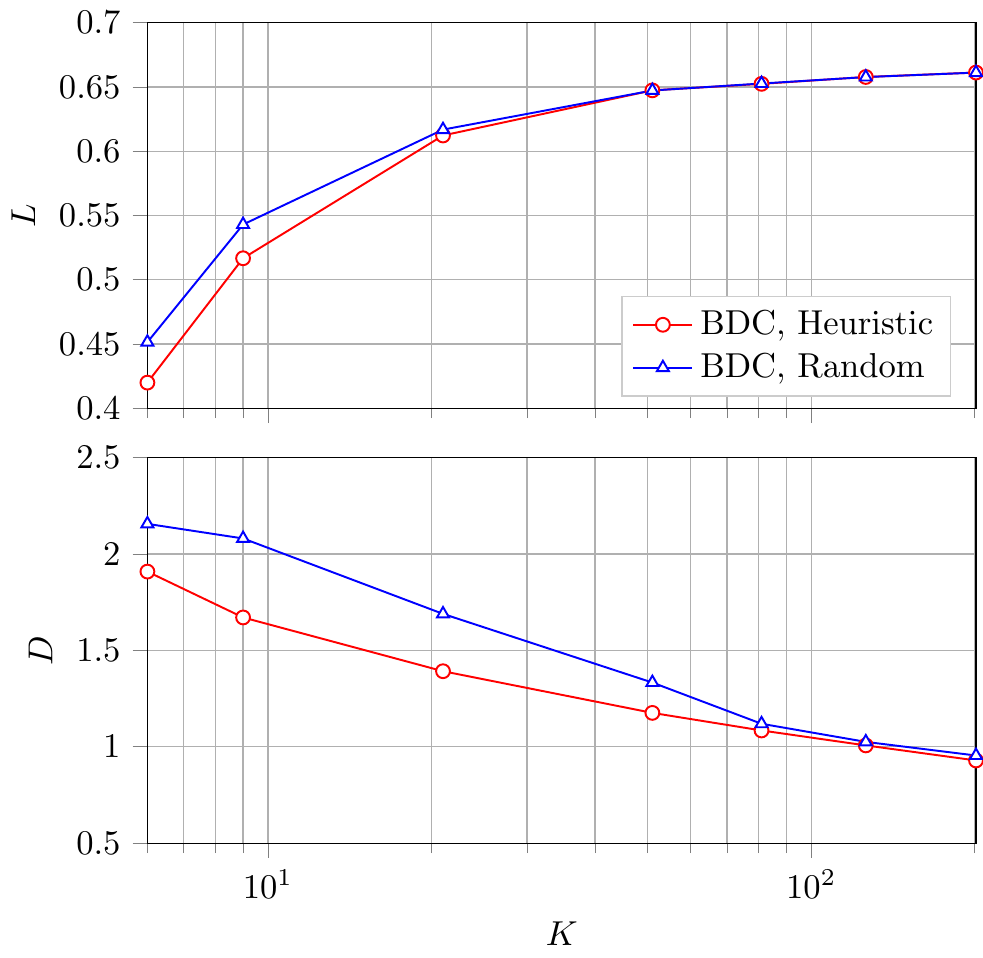}
  }
  \vspace{-4ex}
  \caption{Solver performance as a function of system
    size for $\eta q=2$, $n=m/100$, $\eta m=2000$, code rate
    $m/r=2/3$, and $N=500q$ vectors.}
  \vspace{-3ex}
  \label{fig:solvers_size}
\end{figure}

In Figs.~\ref{fig:solvers_partitions} and \ref{fig:solvers_size}, we plot the
performance of the BDC scheme with assignment $\assign$ given by the
heuristic and the hybrid solver. We also give the average performance
over $100$ random assignments. The vertical dotted line marks the
partitioning limit of Theorem~\ref{pr:performance}. The parameters in
Figs.~\ref{fig:solvers_partitions} and \ref{fig:solvers_size} are identical to those
in Figs.~\ref{fig:numerical_partitions} and \ref{fig:numerical_size}, respectively.

In Fig.~\ref{fig:solvers_partitions}, we plot the performance as a
function of the number of partitions, $T$. For $T$ less than about
$200$, the performance for all solvers is identical.  On the other
hand, for $T>200$ both the computational delay and the communication
load are reduced with $\assign$ from the heuristic solver over the
random assignments (about $5$\% for load and $47$\% for delay at
$T=3000$). A further improvement in communication load can be achieved
using the hybrid solver, but at the expense of a possibly larger
computational delay.

In Fig.~\ref{fig:solvers_size}, we plot the performance as a function of
the number of servers, $K$. The results shown are averages over $1000$
randomly generated realizations of $\G$. For $K=6$, the communication
load of the heuristic solver is about $5\%$ lower than that of the
random assignments, but for $K=201$ the difference is negligible. In
terms of computational delay, the heuristic solver outperforms the
random assignments by about $18$\% and $3$\% for $K=9$ and $K=201$,
respectively. The hybrid solver is too computationally complex for use
with the largest systems considered.

\subsection{Tradeoff Between Communication Load and Computational Delay}

In Fig.~\ref{fig:tradeoff}, we show the tradeoff between communication
load and computational delay. The parameters are $K=14$, $m=50000$
($\pm 3$\% to find valid parameters), $n=500$, $N=840$, and
$\eta=1/2$. Note that the code rate is decreasing toward the bottom of
the plot.  We select the number of partitions $T$ that minimizes the
delay while the load is at most $1$\% or $10$\% higher compared to the
unified scheme. Allowing a $10$\% increased load gives up to about
$7$\% lower delay compared to allowing a $1$\% increase.  For the
topmost data point of the BDC and unified schemes the encoding
complexity dominates, and there is no reason to operate at this
point since both the delay and load can be reduced.  The parameters of
the partitioned LT code-based scheme are $\epsilon_\mathsf{min}=0.3$ and
$\Pft = 10^{-1}$.  For the data point with minimum computational
delay, the LT code-based scheme yields about $15$\% lower delay at the
expense of about a $30$\% higher load compared to the BDC
scheme. Finally, the computational delay of the BDC scheme is between
about $47$\% and $4$\% lower compared to the unified scheme for the
topmost and bottommost data points, respectively.

\subsection{Computational Delay Deadlines}

\begin{figure}[t]
  \centering
  \hspace{3ex} \resizebox{1.05\columnwidth}{!}{
    \includegraphics{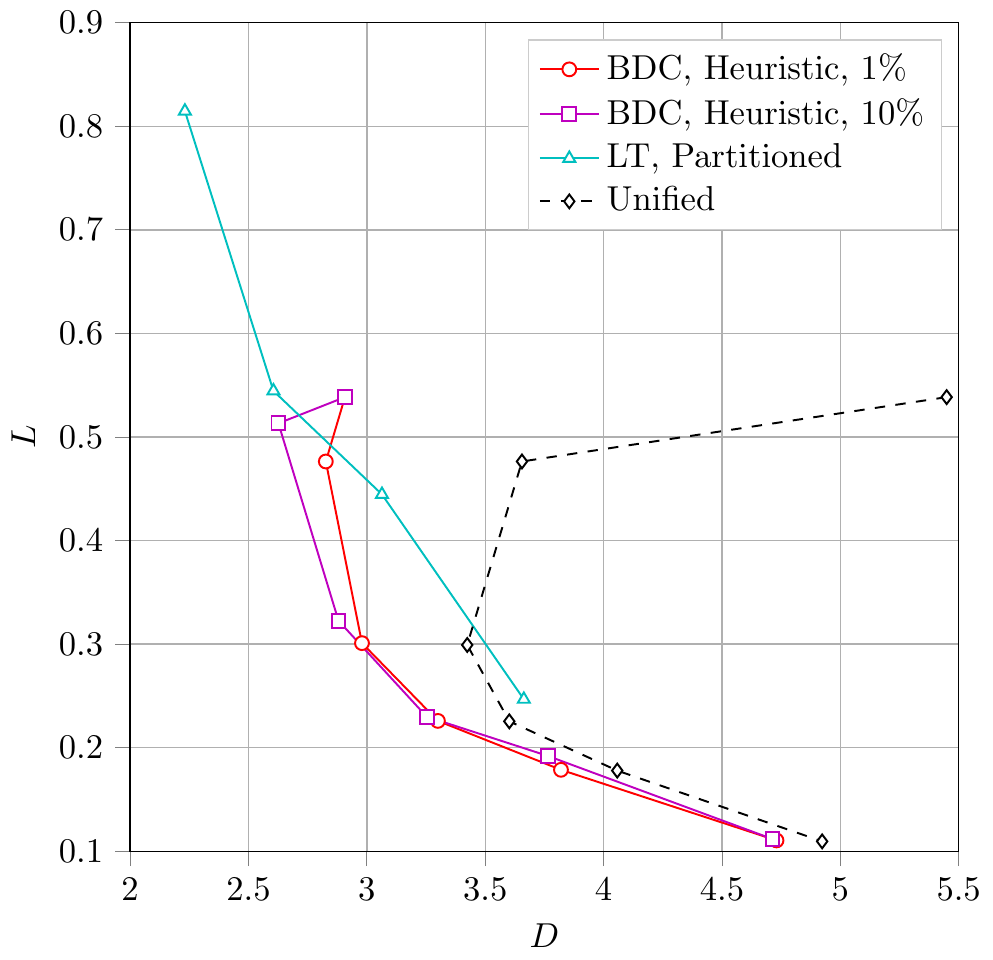}
  }
  \vspace{-4ex}
  \caption{The tradeoff between communication load and computational
    delay for $K=14$, $m = 50000 \pm 3$\%, $n=500$, $N=840$, and
    $\eta=1/2$.}
  \vspace{-2ex}
  \label{fig:tradeoff}
\end{figure}
\begin{figure}[t]
  \hspace{-2ex} \resizebox{1.05\columnwidth}{!}{
    \includegraphics{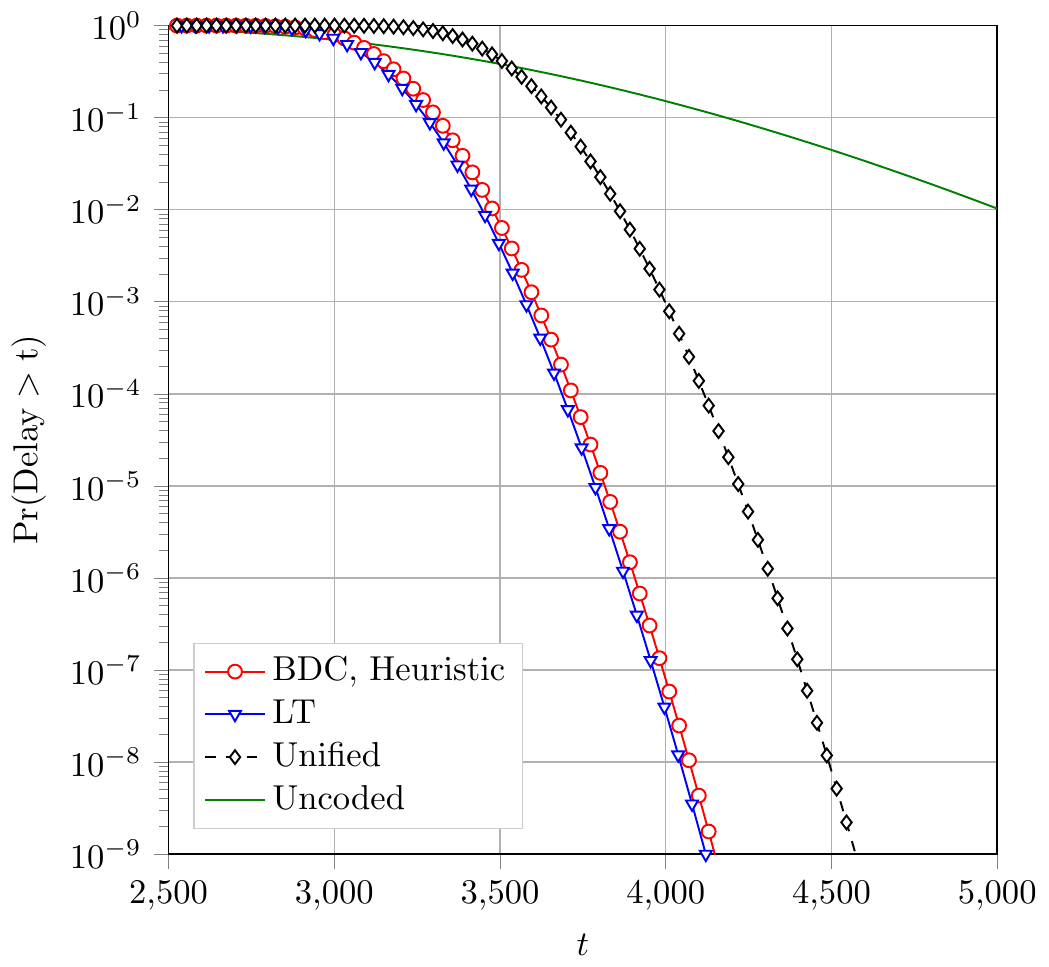}
  }
  \vspace{-3.5ex}
  \caption{The probability of a computation not finishing before a
    deadline $t$ for $K=201$, $q=134$, $m=134000$, $n=1340$, $N=67000$
    vectors, $T=6700$ partitions, code rate $m/r=2/3$,
    $\epsilon_\mathsf{min}=0.335$, and $\Pft=10^{-9}$.}
  \vspace{-3ex}
  \label{fig:deadline}
\end{figure}

In Fig.~\ref{fig:deadline}, we plot the probability of a computation
not finishing before a deadline $t$, i.e., the probability of the
computational delay being larger than $t$. As in \cite{Dutta2017}, we
plot the complement of the CDF of the delay in logarithmic scale. On
the horizontal axis, we show the deadline $t$. The system parameters
are $K=201$, $q=134$, $m=134000$, $n=1340$, $N=67000$ vectors,
$T=6700$ partitions, and code rate $m/r=2/3$. The parameters for the
LT code-based scheme are $\epsilon_\mathsf{min}=0.335$ and
$\Pft = 10^{-9}$. The results are due to simulations. In particular,
we simulate the decoding failure probability of LT codes for various $t$ and
extrapolate from these points under the assumption that the decoding
failure probability is Gamma distributed. The fitted values deviate
negligibly from the simulated values.

When the deadline is $t=3500$, the probability of exceeding the
deadline is about $0.4$ for the unified and uncoded schemes. For the
BDC scheme the probability is only about $7 \cdot 10^{-3}$. The
probability is slightly lower for the LT code-based scheme, about
$4 \cdot 10^{-3}$. If we instead consider a deadline $t=4000$, the
probability of exceeding the deadline is about $10^{-3}$ and $0.15$
for the unified and uncoded schemes, respectively. For the BDC scheme
the probability of exceeding the deadline is about $9\cdot 10^{-8}$,
i.e., $4$ orders of magnitude lower compared to the unified
scheme. The LT code-based scheme further improves the performance with
a probability of exceeding the deadline of about $3\cdot 10^{-8}$. We
remark that for the data point with minimum delay in
Fig.~\ref{fig:tradeoff}, the LT code-based scheme has a significant
advantage over the BDC scheme in terms of meeting a short deadline.

\subsection{Alternative Runtime Distribution}
\label{sec:fixed_tail}

\begin{figure}[t]
  \hspace{-2ex} \resizebox{0.97\columnwidth}{!}{
    \includegraphics{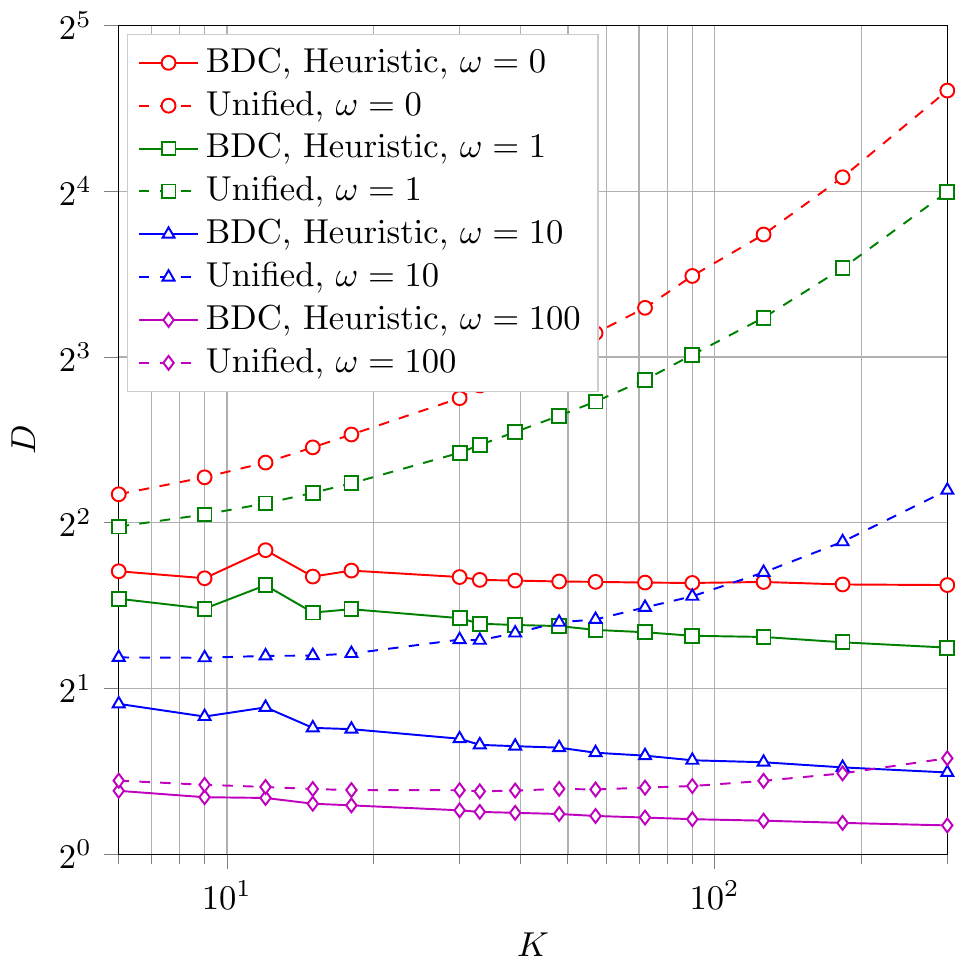}
  }
  \vspace{-1.5ex}
  \caption{Computational delay as a function of system size for
    varying scale of the tail of the runtime distribution. The system
    parameters and communication load are identical to those in
    Fig.~\ref{fig:workload}.}
  \vspace{-3ex}
  \label{fig:altruntime}
\end{figure}

Here, we consider a runtime distribution with CDF
\begin{equation} \notag
  F_H(h; \sigma) = \begin{cases}
    1 - \me^{-\left(h - \sigma \right) / \beta}, & \text{for $h \geq \sigma$} \\
    0, & \text{otherwise}
\end{cases},
\end{equation}
where $\sigma$ is the shift and $\beta$ is a parameter that scales the
tail of the distribution, i.e., it differs from the one considered
previously by that the scale of the tail may be different from the
shift. It is equal to the previously considered distribution if
$\beta=\sigma$. This model has been used to model distributed
computing in, e.g., \cite{Wang2015}. Under this model we assume that
the reduce delay of the uncoded scheme follows the distribution above
with parameters $\beta$ and $\sigma_\mathsf{UC, reduce}=0$ since each
server has to assemble the final output from the intermediate results
regardless coding is used or not. We assume that the encoding delay of
the uncoded scheme is zero. Denote by $\sigma_\mathsf{c}$ the
computational complexity of matrix-vector multiplication for the BDC
and unified schemes. We let $\beta=\omega \sigma_\mathsf{c}$ for
$\omega = 0, 1, 10, 100$. In Fig.~\ref{fig:altruntime}, we plot the
computational delay normalized by that of the uncoded scheme. The
system parameters (and thus also the communication load) are identical
to those in Fig.~\ref{fig:workload}.

We observe the greatest gain of the BDC scheme over the unified scheme
for small $\omega$ since the benefits of straggler coding are small
compared to the added delay due to encoding and decoding, which is
significant for the unified scheme. For larger $\omega$ the benefits
of straggler coding are larger while the delay due to encoding and
decoding remains constant. Hence, the performance of both schemes
converge. However, even for $\omega=100$ the delay of the unified
scheme is about $33$\% higher than that of the BDC scheme for the
largest system considered ($K=300$). We remark that for the example
considered in \cite{Wang2015} the parameters $\beta=\sigma=1$, i.e.,
$\omega=1$, are used.

\section{Conclusion}
\label{sec:conclusion}

We introduced two coding schemes for distributed matrix
multiplication. One is based on partitioning the matrix into
submatrices and encoding each submatrix separately using MDS
codes. The other is based on LT codes. Compared to the earlier scheme
in \cite{Li2016} and to the CMR scheme in \cite{Li2015}, both proposed
schemes yield a significantly lower overall computational delay.  For
instance, for a matrix of size $59800 \times 598$, the BDC scheme
reduces the computational delay by about a factor $20$ over the scheme
in \cite{Li2016} with about a $1$\% increase in communication
load. The LT code-based scheme may reduce the computational delay
further at the expense of a higher communication load. For example,
for a matrix with about $50000$ rows, the computational delay of the
LT code-based scheme is about $15$\% lower than that of the BDC scheme
with a communication load that is about $30\%$ higher. Finally, we
have shown that the proposed coding schemes significantly increase the
probability of a computation finishing within a deadline. The LT
code-based scheme may be the best choice in situations where high
reliability is needed due to its ability to decrease the computational
delay at the expense of the communication load.

\section*{Acknowledgment}
The authors would like to thank Dr.\ Francisco L{\'a}zaro and
Dr.\ Gianluigi Liva for fruitful discussions and insightful comments on
LT codes.

\appendices

\ifCLASSOPTIONcaptionsoff
\newpage
\fi
\balance % balance columns. requires the balance package.
\bibliographystyle{IEEEtran}
%\bibliography{manuscript}{}

\begin{thebibliography}{10}
\providecommand{\url}[1]{#1}
\csname url@samestyle\endcsname
\providecommand{\newblock}{\relax}
\providecommand{\bibinfo}[2]{#2}
\providecommand{\BIBentrySTDinterwordspacing}{\spaceskip=0pt\relax}
\providecommand{\BIBentryALTinterwordstretchfactor}{4}
\providecommand{\BIBentryALTinterwordspacing}{\spaceskip=\fontdimen2\font plus
\BIBentryALTinterwordstretchfactor\fontdimen3\font minus
  \fontdimen4\font\relax}
\providecommand{\BIBforeignlanguage}[2]{{%
\expandafter\ifx\csname l@#1\endcsname\relax
\typeout{** WARNING: IEEEtran.bst: No hyphenation pattern has been}%
\typeout{** loaded for the language `#1'. Using the pattern for}%
\typeout{** the default language instead.}%
\else
\language=\csname l@#1\endcsname
\fi
#2}}
\providecommand{\BIBdecl}{\relax}
\BIBdecl

\bibitem{Barroso2009}
L.~A. Barroso and U.~H{\"o}lzle, \emph{The Datacenter as a Computer: An
  Introduction to the Design of Warehouse-Scale Machines}.\hskip 1em plus 0.5em
  minus 0.4em\relax Morgan \& Claypool Publishers, 2009.

\bibitem{Chen2014}
C.~L.~P. { Chen} and C.-Y. Zhang, ``{Data-intensive applications, challenges,
  techniques and technologies: A survey on big data},'' \emph{Information
  Sciences}, vol. 275, pp. 314--347, Aug. 2014.

\bibitem{Barroso2011}
L.~A. Barroso, ``Warehouse-scale computing: The machinery that runs the
  cloud,'' in \emph{Frontiers of Engineering: Reports on Leading-Edge
  Engineering from the 2010 Symposium}.\hskip 1em plus 0.5em minus 0.4em\relax
  Washington, DC: The National Academies Press, 2011, pp. 15--19.

\bibitem{Dean2004}
J.~Dean and S.~Ghemawat, ``{M}ap{R}educe: Simplified data processing on large
  clusters,'' in \emph{Proc.~Conf.~Symp.~Operating Systems Design \&
  Implementation}, San Francisco, CA, Dec. 2004, p.~10.

\bibitem{Zaharia2016}
M.~Zaharia, R.~S. Xin, P.~Wendell, T.~Das, M.~Armbrust, A.~Dave, X.~Meng,
  J.~Rosen, S.~Venkataraman, M.~J. Franklin, A.~Ghodsi, J.~Gonzalez,
  S.~Shenker, and I.~Stoica, ``Apache {S}park: A unified engine for big data
  processing,'' \emph{Communications of the ACM}, vol.~59, no.~11, pp. 56--65,
  Nov. 2016.

\bibitem{Ranjan2014}
R.~Ranjan, ``Streaming big data processing in datacenter clouds,'' \emph{IEEE
  Cloud Computing}, vol.~1, no.~1, pp. 78--83, May 2014.

\bibitem{Li2015}
S.~Li, M.~A. Maddah-Ali, and A.~S. Avestimehr, ``Coded {M}ap{R}educe,'' in
  \emph{Proc.~Allerton Conf.~Commun., Control, and Computing}, Monticello, IL,
  Sep./Oct. 2015, pp. 964--971.

\bibitem{Lee2017}
K.~Lee, M.~Lam, R.~Pedarsani, D.~Papailiopoulos, and K.~Ramchandran, ``Speeding
  up distributed machine learning using codes,'' \emph{IEEE
  Trans.~Inf.~Theory}, vol.~64, no.~3, pp. 1514--1529, Mar. 2018.

\bibitem{Li2016}
S.~Li, M.~A. Maddah{-}Ali, and A.~S. Avestimehr, ``A unified coding framework
  for distributed computing with straggling servers,'' in \emph{Proc. Work.
  Network Coding and Appl.}, Washington, DC, Dec. 2016.

\bibitem{Ishii2014}
H.~Ishii and R.~Tempo, ``The {P}age{R}ank problem, multiagent consensus, and
  web aggregation: A systems and control viewpoint,'' \emph{IEEE Control
  Systems Mag.}, vol.~34, no.~3, pp. 34--53, Jun. 2014.

\bibitem{Lee2017a}
K.~Lee, C.~Suh, and K.~Ramchandran, ``High-dimensional coded matrix
  multiplication,'' in \emph{Proc.~IEEE Int.~Symp.~Inf.~Theory}, Aachen,
  Germany, Jun. 2017, pp. 2418--2422.

\bibitem{Qian2017}
Q.~Yu, M.~A. Maddah-Ali, and A.~S. Avestimehr, ``Polynomial codes: an optimal
  design for high-dimensional coded matrix multiplication,'' in
  \emph{Proc.~Advances Neural Inf.~Processing Systems}, Long Beach, CA, Dec.
  2017, pp. 4403--4413.

\bibitem{Dutta2016}
S.~Dutta, V.~Cadambe, and P.~Grover, ``{S}hort-{D}ot: Computing large linear
  transforms distributedly using coded short dot products,'' in
  \emph{Proc.~Advances Neural Inf.~Processing Systems}, Barcelona, Spain, Dec.
  2016, pp. 2100--2108.

\bibitem{Reisizadeh2017}
A.~Reisizadeh, S.~Prakash, R.~Pedarsani, and S.~Avestimehr, ``Coded computation
  over heterogeneous clusters,'' in \emph{Proc.~IEEE Int.~Symp.~Inf.~Theory},
  Aachen, Germany, Jun. 2017, pp. 2408--2412.

\bibitem{Severinson2017}
A.~Severinson, A.~{Graell i Amat}, and E.~Rosnes, ``Block-diagonal coding for
  distributed computing with straggling servers,'' in \emph{Proc.~IEEE
  Inf.~Theory Work.}, Kaohsiung, Taiwan, Nov. 2017, pp. 464--468.

\bibitem{Luby2002}
M.~Luby, ``{LT} codes,'' in \emph{Proc.~IEEE Symp. Foundations Computer
  Science}, Vancouver, BC, Canada, Nov. 2002, pp. 271--280.

\bibitem{rfc6330}
M.~Luby, A.~Shokrollahi, M.~Watson, T.~Stockhammer, and L.~Minder, ``{RaptorQ
  Forward Error Correction Scheme for Object Delivery},'' Internet Requests for
  Comments, {RFC Editor}, {RFC} 6330, Aug. 2011.

\bibitem{Edmonds2017}
J.~Edmonds and M.~Luby, ``Erasure codes with a hierarchical bundle structure,''
  \emph{IEEE Trans.~Inf.~Theory}, 2017, to appear.

\bibitem{Verma2015}
A.~Verma, L.~Pedrosa, M.~Korupolu, D.~Oppenheimer, E.~Tune, and J.~Wilkes,
  ``Large-scale cluster management at {Google} with {Borg},'' in
  \emph{Proc.~European Conf.~Computer Systems}, Bordeaux, France, Apr. 2015.

\bibitem{Liang2014}
G.~Liang and U.~C. Kozat, ``{TOFEC}: Achieving optimal throughput-delay
  trade-off of cloud storage using erasure codes,'' in \emph{Proc.~IEEE Conf.
  Computer Commun.}, Toronto, ON, Canada, Apr./May 2014, pp. 826--834.

\bibitem{Arnold2008}
B.~C. Arnold, N.~Balakrishnan, and H.~N. Nagaraja, \emph{A First Course in
  Order Statistics}, 2nd~ed.\hskip 1em plus 0.5em minus 0.4em\relax
  Philadelphia, PA, USA: Society for Industrial and Applied Mathematics, 2008.

\bibitem{Walck2007}
\BIBentryALTinterwordspacing
C.~Walck, ``Hand-book on statistical distributions for experimentalists,''
  Particle Physics Group, University of Stockholm, Sweden, Tech. Rep.
  SUF-PFY/96-01, Sep. 2007. [Online]. Available:
  \url{http://staff.fysik.su.se/~walck/suf9601.pdf}
\BIBentrySTDinterwordspacing

\bibitem{Lin2016}
S.-J. Lin, T.~Y. Al-Naffouri, Y.~S. Han, and W.-H. Chung, ``Novel polynomial
  basis with fast {F}ourier transform and its application to {R}eed-{S}olomon
  erasure codes,'' \emph{IEEE Trans.~Inf.~Theory}, vol.~62, no.~11, pp.
  6284--6299, Nov. 2016.

\bibitem{Garr2013}
G.~Garrammone, ``On decoding complexity of {R}eed-{S}olomon codes on the packet
  erasure channel,'' \emph{IEEE Commun.~Lett.}, vol.~17, no.~4, pp. 773--776,
  Apr. 2013.

\bibitem{Severinson2017code}
\BIBentryALTinterwordspacing
A.~Severinson, ``{Coded Computing Tools},'' Aug. 2018. [Online]. Available:
  \url{https://doi.org/10.5281/zenodo.1400313}
\BIBentrySTDinterwordspacing

\bibitem{Schotsch2013}
B.~Schotsch, G.~Garrammone, and P.~Vary, ``Analysis of {LT} codes over finite
  fields under optimal erasure decoding,'' \emph{IEEE Commun.~Lett.}, vol.~17,
  no.~9, pp. 1826--1829, Sep. 2013.

\bibitem{Wales1997}
D.~J. Wales and J.~P.~K. Doye, ``{Global optimization by basin-hopping and the
  lowest energy structures of Lennard-Jones clusters containing up to 110
  atoms},'' \emph{J. Phys. Chem. A}, vol. 101, no.~28, pp. 5111--5116, Jul.
  1997.

\bibitem{Powell1964}
M.~J.~D. Powell, ``An efficient method for finding the minimum of a function of
  several variables without calculating derivatives,'' \emph{The Computer
  Journal}, vol.~7, no.~2, pp. 155--162, Jan. 1964.

\bibitem{Dutta2017}
S.~Dutta, V.~Cadambe, and P.~Grover, ``Coded convolution for parallel and
  distributed computing within a deadline,'' in \emph{Proc.~IEEE
  Int.~Symp.~Inf.~Theory}, Aachen, Germany, Jun. 2017, pp. 2403--2407.

\bibitem{Wang2015}
D.~Wang, G.~Joshi, and G.~Wornell, ``Using straggler replication to reduce
  latency in large-scale parallel computing,'' \emph{ACM SIGMETRICS Perform.
  Eval. Rev.}, vol.~43, no.~3, pp. 7--11, Dec. 2015.

\end{thebibliography}


\end{document}